\documentclass[openacc]{rsproca_new}%%%%where rsproca is the template name
\usepackage{lipsum}
\usepackage{multicol}
\usepackage{multirow,adjustbox}
\usepackage{float,caption}
\usepackage{graphicx, dblfloatfix,tabularx}
\usepackage{hyperref}
\usepackage{enumitem}

\usepackage[linesnumbered,lined,boxruled,commentsnumbered]{algorithm2e}

\usepackage{amsmath}
\usepackage{algpseudocode}
\newcommand{\pard}[2]{\frac{\partial #1}{\partial #2}}
\newcommand{\hot}[1]{\textcolor{red}{#1}}
\DeclareUnicodeCharacter{2212}{-}

\jname{rsif}
\Journal{J R Soc Interface\ }

\begin{document}

%%%% Article title to be placed here
\title{Parameter selection and optimization of a computational network model of blood flow in single-ventricle patients}

\author{%%%% Author details
Alyssa M Taylor-LaPole$^{1}$, L. Mihaela Paun$^{2}$, Dan Lior$^{3}$, Justin D Weigand$^{3}$, Charles Puelz$^{3}$, Mette S Olufsen$^{1}$}

%%%%%%%%%Insert author address here
\address{\normalsize{$^{1}$Department of Mathematics, North Carolina State University, Raleigh, NC, USA\\
%$^{2}$North Carolina School of Science and Mathematics, Durham, NC, USA\\
$^{2}$School of Mathematics and Statistics, University of Glasgow, Glasgow, UK\\
$^{3}$Division of Cardiology, Department of Pediatrics, Baylor College of Medicine and Texas Children's Hospital, Houston, TX, USA}
}

%%%% Subject entries to be placed here %%%%
\subject{Computational modeling, Parameter inference, Applied mathematics, Single ventricle disease}

%%%% Keyword entries to be placed here %%%%

%%%% Insert corresponding author and its email address}
\corres{Mette S Olufsen\\
\email{msolufse@ncsu.edu}\\
\vspace{0.5 cm}}

%%%% Abstract text to be placed here %%%%%%%%%%%%
\begin{abstract}
%Has to be 200 words or less 
Hypoplastic left heart syndrome (HLHS) is a congenital heart disease responsible for 23\% of infant cardiac deaths each year \hot{in the United States}. HLHS patients are born with an underdeveloped left heart, requiring several surgeries to reconstruct the aorta and create a single ventricle circuit known as the Fontan circulation. While survival into early adulthood is becoming more common, Fontan patients suffer from reduced cardiac output, putting them at risk for a multitude of complications. These patients are monitored using chest and neck MRI imaging, but these scans do not capture energy loss, pressure, wave intensity, or hemodynamics beyond the imaged region. This study develops a framework for predicting these missing features by combining imaging data and computational fluid dynamics (CFD) models. Predicted features from models of HLHS patients are compared to those from control patients with a double outlet right ventricle (DORV). We \hot{infer patient-specific parameters through the proposed framework}. In the calibrated model, we predict pressure, flow, wave-intensity (WI), and wall shear stress (WSS). \hot{Results reveal that HLHS patients have lower compliance than DORV patients}, resulting in lower WSS and higher WI in the ascending aorta and increased WSS and decreased WI in the descending aorta.
\end{abstract}
%%%%%%%%%%%%%%%%%%%%%%%%%%%

%%%%%%%%%% Insert the texts which can accomdate on firstpage in the tag "fmtext" %%%%%
\begin{fmtext}
\vspace*{-0.5cm}
 
\end{fmtext}
%%%%%%%%%%%%%%% End of first page %%%%%%%%%%%%%%%%%%%%%

\maketitle

\begin{multicols}{2}
\section{Introduction}
\noindent Hypoplastic left heart syndrome (HLHS) is a congenital heart disease responsible for 23$\%$ of infant cardiac deaths and up to 9$\%$ of congenital heart disease cases each year \hot{in the United States} \cite{Fruitman2000}. The disease arises in infants with an underdeveloped left heart, preventing adequate transport of oxygenated blood to the systemic circulation \cite{Tworetzky2001}. Characteristics of HLHS include an underdeveloped left ventricle and ascending aorta, as well as small or missing \hot{aortic} and mitral valves. Three surgeries are performed over the patients' first 2-3 years of life, resulting in a the Fontan circuit (Figure~\ref{fig:phys}), a fully functioning, univentricular circulation. 
The first surgical stage in the creation of the Fontan circulation involves moving the aorta from the left to the right ventricle and widening it with tissue from the pulmonary artery. This surgically modified aorta is hereafter referred to as the ``reconstructed'' aorta. Next, the venae cavae are removed from the right atrium and attached to the pulmonary artery. As a result, flow to the pulmonary circulation is achieved by passive transport through the systemic veins \cite{Fontan1971,Gobergs2016,Mahle1998}. The Fontan circuit has near-normal arterial oxygen saturation. However, the lack of a pump pushing blood into the pulmonary vasculature causes a bottleneck effect, increasing pulmonary impedance and decreasing venous return to the heart \cite{Gewillig2016}. The single-pump system degenerates over time due to vascular remodeling accentuated by the system's attempt to compensate for reduced cardiac output \cite{Mahle1998,Voges2010,Gewillig2016}. Clinical implications of the Fontan physiology include reduced cerebral and gut perfusion, increasing the risk of stroke \cite{Saiki2014} and the development of Fontan-associated liver disease (FALD) \cite{Gordon2019,Navaratnam2016}. The study by Saiki \textit{et al.} \cite{Saiki2014} shows that increased stiffness of the aorta, head, and neck vessels in HLHS patients with reconstructed aortas reduces blood flow to the brain. This reduced flow, combined with increased arterial stiffness, is associated with ischemic stroke \cite{Tsivgoulis06}. Regarding FALD, the single pump system decreases both supply and drainage of blood in the liver. Hypertension within the liver vasculature typically occurs 5-10 years after Fontan surgery \cite{Gordon2019}. Reduced cardiac output and passive venous flow lead to liver fibrosis, a characteristic of FALD. With a heart transplant, FALD is reversible if caught early. Advanced FALD is irreversible and can lead to organ failure \cite{Gordon2019}. 

Another single ventricle pathology is the double outlet right ventricle (DORV), also treated by creating a Fontan circulation. DORV patients, also have a non-functioning left ventricle, but are born with a fully functioning aorta attached to the right ventricle. This physiology makes aortic reconstruction unnecessary \cite{Russo08}. Figure ~\ref{fig:phys} shows a healthy (a) and a single ventricle (b) heart and circulation. The gray patch on the aorta indicates the reconstructed portion, required in HLHS but not DORV patients. Studies \cite{Rutka2018,Sano2022} suggest that DORV patients undergo less vascular remodeling and, therefore, have better cardiac function. Rutka \textit{et al.} \cite{Rutka2018} found that Fontan patients with reconstructed aortas had poorer long-term survival, and Sano \textit{et al.} \cite{Sano2022} noted that these patients have an increased need for surgical intervention and increased mortality rates.
\noindent\begin{minipage}{\columnwidth}
\captionsetup{type=figure}
\centering
\includegraphics[width=1.0\textwidth]{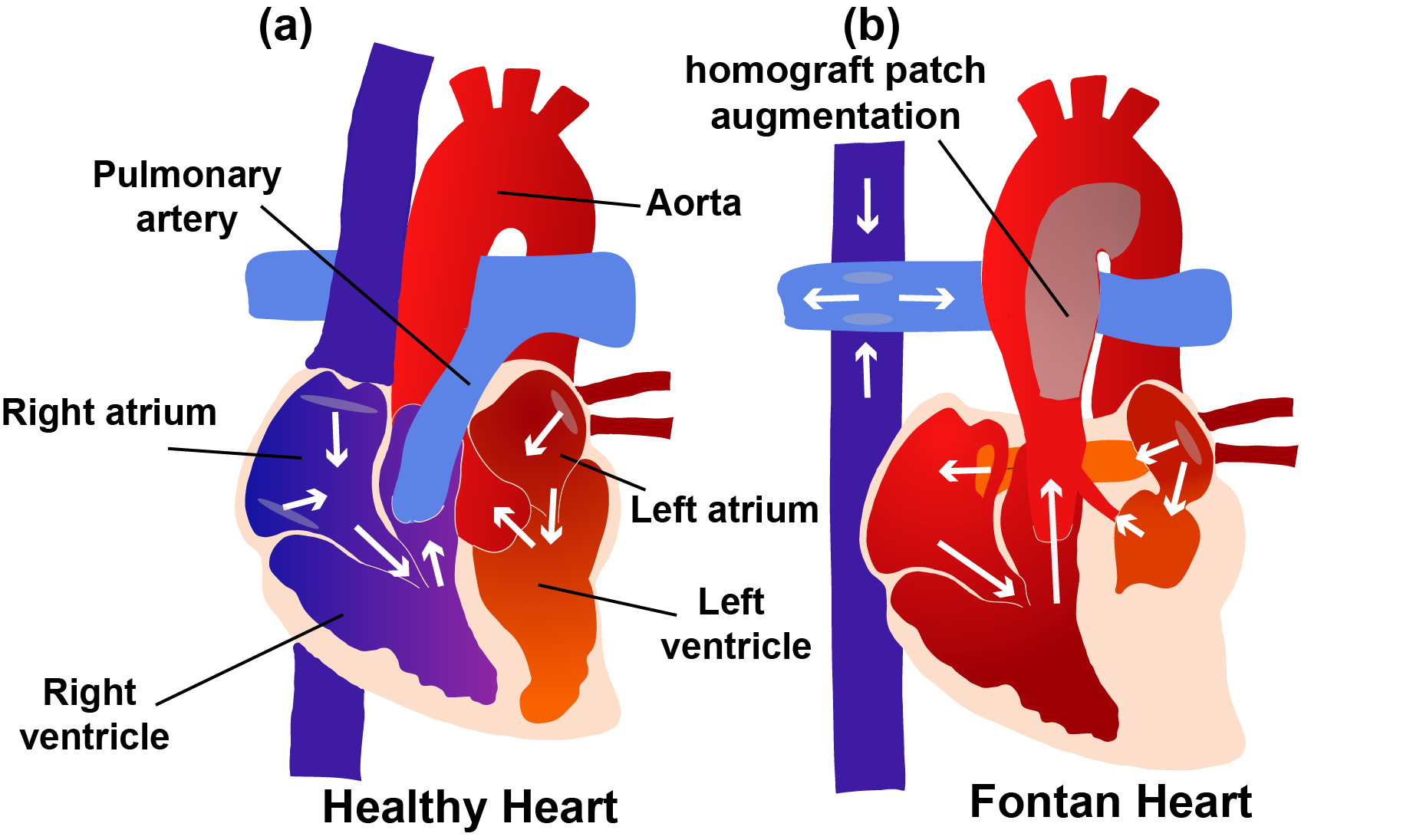}
    \caption{Comparison of healthy (a) and single ventricle (b) hearts and vessels. The homographic patch augmentation generated on HLHS patients’ ascending aorta is highlighted in gray in panel (b). DORV patients have a Fontan circulation but do not need aortic reconstruction. \hot{The white arrows indicate the direction of blood flow.}}
\label{fig:phys}
\end{minipage}

Single ventricle patients are monitored throughout life to assess the function of their ventricle pump \cite{Fruitman2000}. Typically, this is done via time-resolved magnetic resonance imaging (4D-MRI) \cite{Stankovic2014,Markl2012,Soulat2020} and magnetic resonance angiography (MRA) of vessels in the neck and chest. 4D-MRI provides time-resolved blood velocity fields in large vessels, while MRA provides a high-fidelity image of the vessel anatomy. These imaging sequences do not measure energy loss, blood pressure, wave intensity, or hemodynamics outside of the imaged region. Additional hemodynamic information can be predicted by combining imaging data with computational fluid dynamics (CFD) modeling \cite{LaPole2022,Mittal2016,Young2008,Bordones2018,Chinnaiyan2017}.
\begin{figure*}
    \centering
    \includegraphics[width=1.0\linewidth]{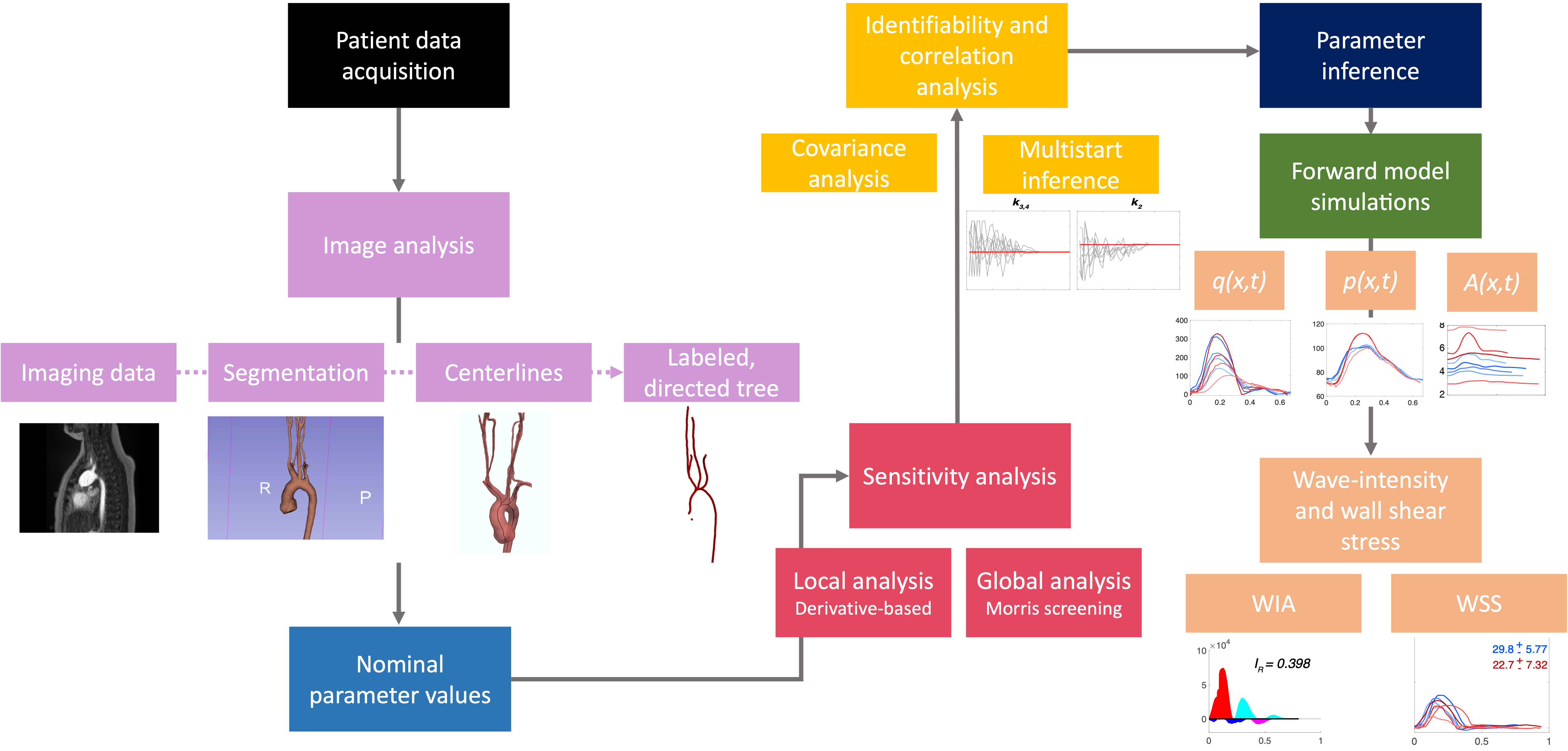}
    \caption{Workflow of the methods implemented in this study. We begin with patient data acquisition, consisting of biometrics, 4D-MRIs, and MRAs. We then perform image analysis to obtain patient geometries. We set nominal parameter values for each patient and performed sensitivity analysis. Once we have determined influential parameters, we use covariance analysis to choose a subset of identifiable parameters estimated using multistart parameter inference. Estimated parameters are used for forward model simulations for each patient.}
    \label{fig:workflow}
\end{figure*}

Most computational work on the Fontan circulation focuses on venous blood flow. This includes studies investigating the hemodynamics and function of the total cavopulmonary connection \cite{Marsden2007,Marsden2009,Ahmed2021}. However, computational studies assessing the Fontan circuit from the systemic arterial perspective are scarce. A few have examined systemic arterial hemodynamics in individual patients. This includes the study by Taylor-LaPole \textit{et al.} \cite{LaPole2022} using a one-dimensional (1D) CFD model to compare DORV and HLHS flow and pressure wave propagation in the cerebral and gut vasculature under rest and exercise conditions. Their study is promising, but results are calibrated manually and only include a single DORV and HLHS patient pair. The study by Puelz \textit{et al.} \cite{Puelz2017} used a 1D CFD model to explore the effects of two different Fontan modifications (fenestration versus hepatic vein exclusion) on blood flow to the liver and intestines. Their study compared model predictions to ranges of clinical data obtained from literature. Still, they lacked a systematic methodology to determine model parameters. These studies successfully built patient-specific networks and predict hemodynamics outside the imaged region, demonstrating the importance of calibrating models to data but lack a systematic methodology to determine model parameters. 

This limitation is addressed by combining imaging and hemodynamic data with a 1D CFD model. The model is calibrated to patient-specific data using sensitivity analysis and parameter inference, an improvement over previous studies that rely on manual tuning of parameters. This approach presented here is  used to predict hemodynamics in four matched HLHS and DORV patient pairs.

Calibration of the model is crucial, especially when it is used for predictions in a clinical setting. For this purpose, an identifiable subset of parameters is needed. Calibrating the model by estimating identifiable parameters provides unique patient-specific biomarkers, which can be compared between patients. Two steps are used to determine an identifiable parameter subset: sensitivity analyses (local and global); determining the effect of a given parameter on a specified quantity of interest \cite{Colebank_sens}, and subset selection, characterising interactions among parameters \cite{Karau2011,Olsen2013}. Several recent studies use these methodologies. Schiavazzi \textit{et al.} \cite{Schiavazzi2017} combine local and global sensitivity analyses to find a set of parameters that fit a zero-dimensional model to data from single-ventricle patients. This study demonstrates that using both techniques lead to consistent and accurate subset selection. The study by Colebank \textit{et al.} \cite{Colebank_sens} uses local and global sensitivity analyses to determine the most influential parameters. They combine results with covariance analysis to quantify parameter interactions and further reduce the parameter subset to be inferred. The study calibrates the model using experimental data, but they only have pressure data at one location. Colunga \& Colebank \textit{et al.} \cite{Colunga23} used local and global sensitivity analyses for subset selection. Their study incorporated multi-start inference to ensure identifiability and low coefficient of variation for all parameters. 

Inspired by these studies, we use local and global sensitivity analyses to determine parameter influence and covariance analysis to identify correlated parameters. We then use multistart inference to estimate the identifiable parameters. Once calibrated, we predict hemodynamic quantities, including pressure, flow, wave intensity, and wall shear stress to compare cardiac function between HLHS and DORV patients. \hot{This study aims to create an efficient and accurate parameter inference pipeline for these single-ventricle patients. Parameter inference and model calibration are crucial to ensure accurate hemodynamic insights that clinicians for each patient will investigate.} To our knowledge, this is the first study that uses multiple data sets to calibrate a 1D, patient-specific CFD model for single-ventricle patients.

\section{Methods}
Medical imaging and hemodynamic data are combined to construct a patient-specific 1D arterial network model. Magnetic resonance angiography (MRA) images \cite{Stankovic2014} are segmented, creating a 3D representation of vessels within the imaged region. For each HLHS and DORV patient, we extract vessel dimensions (radius and length) and vessel connectivity. The MRA images are registered to the 4D-MRI images, allowing for the extraction of flow waveforms in the aorta, head, and neck vessels using the high-fidelity MRA anatomy. \hot{The network includes vessels in the imaged region extending terminal vessels and including one more bifurcation.} This ensures sufficient resolution of wave reflections. We solve a 1D CFD model in this network and use sensitivity analyses and subset selection to determine a subset of identifiable parameters. Parameter inference uses a gradient-based optimization scheme to minimize the residual sum of squares (RSS). A full workflow of our methods can be seen in Figure \ref{fig:workflow}.

\begin{table*}[b]
\centering
\caption{Patient data noting age, height, weight, gender, systolic and diastolic pressures, cardiac output (CO), and cardiac cycle length ($T$).}
    \begin{tabular}{c c r r r r r r r}
    \hline\noalign{\smallskip}
      \hot{\textbf{Pair}} & \hot{\textbf{Patient}} & \hot{\textbf{Age}} & \hot{\textbf{Height (cm)}} & \hot{\textbf{Weight (kg)}} & \hot{\textbf{Gender}} & \hot{\textbf{Pressure (mmHg)}} & \hot{\textbf{CO (L/min)}} & \hot{$\mathbf{T}$} \hot{\textbf{(s)}}\\
      \hline\noalign{\smallskip}
      \multirow{2}{3em}{\hot{\textbf{Pair 1}}} & \hot{DORV 1} & \hot{16} & \hot{171.7} & \hot{52.6} & \hot{M} & \hot{105/57} & \hot{4.0} & \hot{0.802}\\
      & \hot{HLHS 1} &  \hot{18} & \hot{159.0} & \hot{57.5} & \hot{M} & \hot{129/72} &  \hot{5.4} & \hot{0.942}\\
      \hline\noalign{\smallskip}
      \multirow{2}{3em}{\hot{\textbf{Pair 2}}} & \hot{DORV 2} & \hot{12} & \hot{154.3} & \hot{59.6} & \hot{F} & \hot{110/67} & \hot{4.0} & \hot{0.658}\\
      & \hot{HLHS 2} & \hot{11} & \hot{151.4} & \hot{62.0} & \hot{M} & \hot{116/65} & \hot{3.9} & \hot{0.615}\\
      \hline\noalign{\smallskip}
      \multirow{2}{3em}{\hot{\textbf{Pair 3}}} & \hot{DORV 3} & \hot{11} & \hot{136.4} & \hot{32.7} & \hot{M} & \hot{119/60} & \hot{3.3} & \hot{0.774}\\
      & \hot{HLHS 3} & \hot{13} & \hot{148.0} & \hot{38.1} & \hot{M} & \hot{99/53} & \hot{2.8} & \hot{0.933}\\
      \hline\noalign{\smallskip}
      \multirow{2}{3em}{\hot{\textbf{Pair 4}}} & \hot{DORV 4} & \hot{12} & \hot{148.5} &\hot{ 42.3} & \hot{F} & \hot{112/6} & \hot{3.7} & \hot{0.598}\\
      & \hot{HLHS 4} & \hot{14} & \hot{163.0} & \hot{53.7} & \hot{M} & \hot{116/60} & \hot{3.2} & \hot{0.605}\\
      \hline\noalign{\smallskip}
    \end{tabular}
    \label{tab:PatientData}
\end{table*}

\subsection{Data acquisition}
\label{sec:Data}
Data include four pairs of age- and sized-matched HLHS and DORV patients seen at Texas Children's Hospital, Houston, TX. Data collection was approved by the Baylor College of Medicine Institutional Review Board (H-46224: “Four-Dimensional Flow Cardiovascular Magnetic Resonance for the Assessment of Aortic Arch Properties in Single Ventricle Patients”). \hot{To isolate the differences in flow properties based on native versus reconstructed aortas, single ventricle morphologies were selected to have the same dominant single right ventricle. The DORV group includes patients with a single ventricle anatomy with right ventricular morphology and a native arch. The HLHS group includes patients with a single ventricle anatomy with right ventricular morphology and reconstructed aortas. Each cohort have similar atrioventricular valve regurgitation, achieved total cavopulmonary anastomosis (non-fenestrated), and preserved right ventricular systolic function. Exclusion criteria include patients with aortic surgery beyond initial Norwood, significant collateral burden $<$40\%, systemic hypertension, and those with heart failure. Patients were matched across cohorts (DORV to HLHS) based on patient body surface area and age.}

\vspace{0.2 cm}

\noindent\textbf{Imaging and pressure data.} Cuff pressures are measured in the supine position. Images are acquired on a 1.5 T Siemens Aera magnet. MRA images are acquired in the sagittal plane and contain a reconstructed voxel size of 1.2 mm$^3$. 4D-MRI images are acquired during free-breathing with a slice thickness of 1-2.5 mm and encoding velocity 10\% larger than the highest anticipated velocity in the aorta. The imaged region contains the ascending aorta, aortic arch, descending thoracic aorta, and innominate artery vessels (Figure \ref{fig:network}).

Patient characteristics, including age (years), height (cm), sex (m/f), weight (kg), systolic and diastolic blood pressure (mmHg, measured with a sphygmomanometer), and cardiac output (L/min), are listed in Table \ref{tab:PatientData}. 

\vspace{0.2 cm}

\noindent\textbf{3D rendering.} 3D rendered volumes are generated from the MRA and 4D-MRI images but vessel dimensions are only extracted from the MRA image. The aorta, head, and neck vessels are segmented using 3D Slicer, an open-source software from Kitware \cite{Federov2012,Kikinis14}. Using thresholding, cutting, and islanding (3D Slicer tools), we create a 3D rendered volume. The threshold for the 4D-MRIs is 0.51-1.00 Hounsfield Units (HU), and the threshold for MRAs is 430-1262 HU. The rendered volumes are saved to a stereolithography (STL) file,  imported into Paraview \cite{Utkarsh15}, and converted to a VTK polygonal data file. 

\vspace{0.2 cm}

\noindent\textbf{Centerlines.} \hot{Using the Vascular Modeling Toolkit (VMTK), centerlines are generated from maximally inscribed spheres (Figure \ref{fig:segment})}. At each point along the centerline, the vessel radius is obtained from the associated sphere \cite{Antiga2008}. Users manually select an inlet and outlet points. VMTK recursively generates centerlines, beginning at the outlets and determining pathways to the inlet. Junctions are defined as the points where two centerlines intersect. This procedure does not place the junction point at the barycenter for most vessels, especially those branching from the aorta. To correct this, we use an in-house algorithm \cite{BartoloLaPole2023} detailed in the supplement.

\noindent\begin{minipage}{\columnwidth}
\captionsetup{type=figure}
\centering
\includegraphics[width=0.8\textwidth]{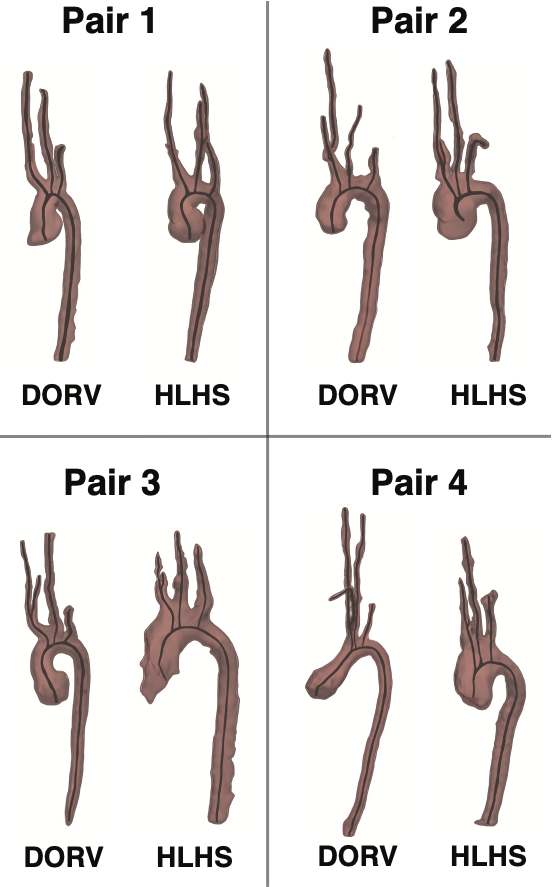}
    \caption{Segmentations and centerlines for each patient shown in their respective pairs.}
    \label{fig:segment}
\end{minipage}

\begin{figure*}
    \centering
    \includegraphics[width=0.6\textwidth]{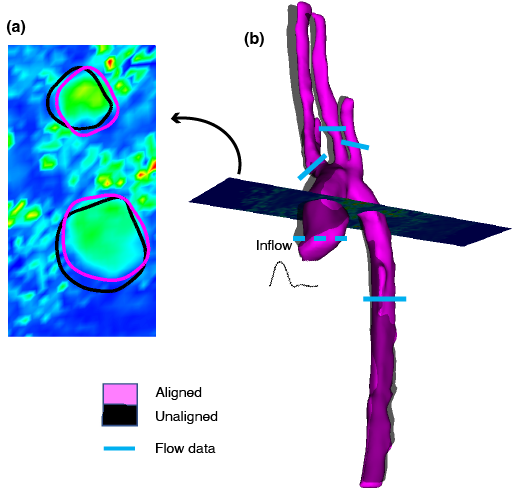}
    \caption{: Before and after registering the MRA to the 4D-MRI. (a) shows velocities from the 4D-MRI in the unaligned (black circles) versus the aligned (pink circles) network. (b) shows aligned (pink) MRA segmentation transposed on the unaligned segmentation (black). The velocities in (a) correspond to the slice through the segmentations in (b). Blue lines denote where flow data was obtained from each patient. The dashed blue line shows where the inflow profile was obtained.}
    \label{fig:registration}
\end{figure*}

\noindent\textbf{Image registration.} Volumetric flow waveforms   are extracted from the 4D-MRI images. The 4D-MRI sequence measures the time-resolved blood velocity field at each voxel in the imaged region with a sampling frequency 19-28 times over the course of a cardiac cycle. 4D-MRIs provide vascular geometry, but only at certain phases of the cardiac cycle and with low spatial resolution. Hence, both an MRA, for the anatomy, and a 4D-MRI, for the velocity, are acquired for each patient. These two images are obtained in the same session, but shifting and deformation may occur between scans caused by patient movement or misalignment within the respiratory cycles. Without correction, this can lead to inaccurate blood flow and volume estimates. To compensate, the MRA and 4D-MRI are aligned via an image registration procedure (shown in Figure \ref{fig:registration}) \cite{Lior2023}. Details description is given in the supplement. 

\vspace{0.2 cm}

\noindent\textbf{Volumetric flow data.} Flow waveforms are extracted from the 4D-MRI image using the aligned MRA segmentation. See supplement for detailed description. 

The 4D-MRI velocity field is generally not divergence-free, in part because of noise in the measurements and averaging performed over several cardiac cycles. Thus, the sum of the flows in the branching vessels does not equal the flow in the ascending aorta. The 1D CFD model assumes that blood is incompressible and flow through vessel junctions is conserved. To ensure convergence of the optimization, we impose volume conservation by scaling the flow waveforms extracted from the 4D-MRI images. The scaling uses a linear system to calculate the smallest possible scale factors needed to enforce mass conservation. The cardiac output of the ascending aorta is held fixed, and flows in the peripheral branching vessels are scaled by
\begin{equation}
    q_{\text{inlet}}=\sum_{i=1}^4\left(1+\alpha_i\right)q_i.
\end{equation}

\noindent The term $q_{\text{inlet}}$ is the flow through the ascending aorta, $\alpha_i$ is the scaling factor for the $i^{\text{th}}$ vessel, and $q_i$ is flow in the $i^{\text{th}}$ vessel. \hot{A linear system for the scaling factors $\alpha_i$ is constructed from the constrained optimization problem, minimizing}
\begin{equation}
    \hot{f=\lambda \left(\sum_{i=1}^n \left(\left(1+\alpha_i\right)q_i\right)-q_{\text{inlet}}\right) + \frac{1}{2n}\sum_{i=1}^{n} \alpha_i^2,}
\end{equation}
\hot{where $n=4$. See supplement for a complete description.} It should be noted that the flow waveforms are determined by extracting five flow waveforms from the 4D-MRI image that lie in close proximity on the centerline and are within a straight section of the vessel. These waveforms are averaged to create a representative flow waveform.
\noindent\begin{minipage}{\columnwidth}
\captionsetup{type=figure}
\centering\includegraphics[width=1.0\textwidth]{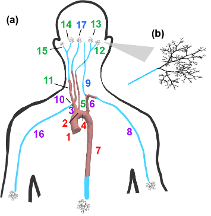}
    \caption{Network used in the fluids model. (a) Arteries extracted from the MRA images (dark red) and extended vessels outside of the imaged region (cyan). Small vessel networks (black), represented by structured trees are attached to the end of each large terminal vessel. (b) Example structured tree including vessels down to the arteriolar level. Vessel numbers correspond to those given in Table \ref{table:Dim}. \hot{The vessel number colors indicate the parameter group used for parameter inference described in Section  \ref{sec:summary}. Red vessel numbers represent group 1, purple group 2, green group 3, and blue group 4.}}
    \label{fig:network}
\end{minipage}

\subsection{Network generation}
\label{sec:Network}
A labeled directed graph is generated for each patient using in-house algorithms \cite{Colebank19,Colebank21}, extracting vessel radii and lengths from the VMTK output. The centerlines are defined as edges, containing $x,y,z$ coordinates along the centerline, and junctions as nodes. Vessel radii are specified at each point along an edge from the radius of the maximally inscribed sphere. The vessel length is calculated as the sum of the Euclidean distances between the $x,y, \text{ and } z$ coordinates. Nodes shared between edges spanning more than one vessel are called ``junction nodes.'' Terminal vessels are defined as those with no further branching. A connectivity matrix is used to specify how vessels connect to each other in the network. For example, in Figure \ref{fig:network}, vessel 1 (the ascending aorta) forms a junction with vessel 2 (the aortic arch) and vessel 3 (the innominate artery). 

The generated network is a tree. The tree has a natural direction since flow moves from the heart towards the peripheral vasculature. The network is represented by a labeled directed graph, where labels include the vessel radii and lengths. The directed graph generated from the images is adjusted to move junctions to the barycenter and determine representative vessel radii. This adjustment is done using an in-house recursive algorithm \cite{BartoloLaPole2023}.

\hot{The network is extended beyond the imaged region, extending terminal vessels to include one additional bifurcation to capture wave reflections in model predictions \cite{LaPole2022}.} In total, $17$ vessels are included in the network, with 9 created from the imaged region. Allometric scaling \cite{Pennati2000} is performed on vessel dimensions outside the imaged region. We scale values obtained from the literature \cite{Olufsen2001} as
\begin{equation}
    L_2 = L_1\left(\frac{W_1}{W_2}\right)^\alpha,
\end{equation}
where $\alpha = 0.35$, $W_1$ (kg) and $L_1$ (cm) are literature bodyweight and vessel length values, $W_2$ (kg) is the bodyweight of the patient, and $L_2$ (cm) is the unknown vessel length. Vessel radii not specified in the imaged region are calculated similarly. 

At terminal ends of the large vessels, asymmetric structured trees are used for outlet boundary conditions for the fluids model. \hot{The structured trees represent arterial branching down to the arteriolar level (Figure \ref{fig:network})}.

\subsection{Fluid dynamics}
\label{sec:Fluids}
A 1D fluid dynamics model derived from the Navier-Stokes equations is used to compute pressure $p(x,t)$ (mmHg), flow $q(x,t)$ (mL/s), and cross-sectional area $A(x,t)$ (cm$^2$). In the large vessels, the equations are solved explicitly, and in the small vessels, a wave equation is solved in the frequency domain to predict impedance at the root of the small vessels.

\vspace{0.2 cm}

\noindent\textbf{Large vessels.} We assume the blood to be incompressible, Newtonian, and homogeneous with axially symmetric flow, and the vessels to be long and thin.  Under these assumptions, flow, pressure, and vessel cross-sectional area satisfy mass conservation and momentum balance equations of the form
\begin{align}
    \pard{A}{t}+\pard{q}{x} &= 0 \label{eq:mass},\\
    \pard{q}{t}+\pard{}{x}\left(\frac{q^2}{A}\right)+\frac{A}{\rho}\pard{p}{x} &= -\frac{2\pi\nu R}{\delta}\frac{q}{A},
    \label{eq:momentum}
\end{align}
\hot{where $0\leq x\leq L$ (cm) is the axial position along the vessel, $\mu$ (g/cm-s) is the viscosity, $\rho$ (g/cm$^3$) is blood density, $\nu = \mu/\rho$ (cm$^2$/s) is the kinematic viscosity, $R$ (cm) is the radius, and $t$ (s) is time}. These equations rely on the  assumption of a Stokes boundary layer 
\begin{equation}
\label{eq:stokes}
u_x(r,x,t) = \left\{
  \begin{array}{lr} 
      \displaystyle \bar{u}_x, \hspace{1cm} & r<R-\delta, \\
       \displaystyle\bar{u}_x\frac{(R-r)}{\delta}, \hspace{1cm} & R-\delta<r\leq R.
      \end{array}
\right.
\end{equation}
Here $\delta = \sqrt{\nu T/2\pi}$ (cm) is the boundary layer thickness, $T$ (s) is the duration of the cardiac cycle, $u_x$ is the axial velocity, and $\bar{u}_x$ is the axial velocity outside of the boundary layer. 

To close our system of equations, we define a pressure-area relationship of the form
\begin{align}
    p(x,t) -p_0 &= \frac{4}{3}\frac{Eh}{r_0}\left(1-\sqrt{\frac{A_0}{A}}\right), \label{eq:wallmodel}\\
    \frac{Eh}{r_0} &= k_1\exp(-k_2r_0)+k_3 \label{eq:stiff},
\end{align}
where $E$ (g/cm/s$^2$) is Young's modulus, $h$ (cm) is the vessel wall thickness, $p_0$ (g/cm/s$^2$) is a reference pressure, $r_0$ (cm) is the inlet radius, and $A_0$ (cm$^2$) is the cross-sectional area when $p(x,t) = p_0$. \hot{The area $A_0=\pi r_0^2$  is determined from the vessel network as described in Section \ref{sec:Network}.}

The system \eqref{eq:mass}-\eqref{eq:momentum} is hyperbolic with Riemann invariants propagating in opposite directions. Therefore, boundary conditions are required at the inlet and outlet of each vessel. At the ascending aorta, corresponding to the root of the labeled tree, an inlet flow waveform is imposed using the patient-specific flow waveform extracted from the 4D-MRI data.  At junctions, we impose mass conservation and pressure continuity
\begin{equation}
    q_p = q_{d1}+q_{d2}, \hspace{0.3cm}
    p_p = p_{d1} = p_{d2}.
\end{equation}
The subscripts $p$ and $d1,2$ refer to the parent and daughter vessels, respectively. Outlet boundary conditions are imposed by coupling the terminals of the large vessels to an asymmetrical structured tree model (described below) representing the small vessels \cite{Olufsen2001}. The model equations are nondimensionalized and solved using the explicit two-step Lax-Wendroff method \cite{Olufsen2001,LaPole2022,BartoloLaPole2023,Colebank19,Colebank21}.

\vspace{0.2 cm}

\noindent\textbf{Small vessels.} In the small vessels, viscous forces dominate; therefore, we neglect the nonlinear inertial terms. As a result, equations (\ref{eq:mass}) and (\ref{eq:momentum}) are linearized and reduced assuming periodicity of solutions, resulting in
\begin{align}
    i\omega Q+\frac{A_0\left(1-F_J\right)}{\rho}\pard{P}{x} &=0, &F_J = \frac{2J_1(\omega_0)}{\omega_0J_0(\omega_0)} \label{eq:linmass}\\
    i\omega CP+\pard{Q}{x} &=0, &C = \pard{p}{A}\approx\frac{3}{2}\frac{r_0}{Eh}, \label{eq:linmom}
\end{align}
where $J_0$ and $J_1$ are the zeroth and first order Bessel functions, $\omega_0^2=i^3r_0\omega/\nu$ is the Womersely number, and $C$ denotes the vessel compliance. The term $Eh/r_0$ has the same form as (\ref{eq:stiff}) but with small vessel stiffness parameters, $ks_1,ks_2, \text{ and } ks_3$. Analytical solutions for equations (\ref{eq:linmass}) and (\ref{eq:linmom}) are
\begin{align*}
    Q(x,\omega) &= a\cos\left(\frac{\omega x}{c}\right)+b\sin\left(\frac{\omega x}{c}\right),\\
    P(x,\omega) &= \frac{i}{g_\omega}\left(-a\sin\left(\frac{\omega x}{c}\right)+b\cos\left(\frac{\omega x}{c}\right)\right),
\end{align*}
where $a,b$ are integration constants and $g_\omega = \sqrt{CA_0(1-F_J)/\rho}$ \cite{BartoloLaPole2023,Olufsen2001,Colebank21,LaPole2022}.  From this equation, it is possible to determine impedance at the beginning of each vessel as a function of the impedance at the end of the vessel, 
\begin{equation}
Z(0,\omega)=\frac{ig^{-1}\sin\left(\frac{\omega L}{c}\right)+Z(L,\omega)\cos\left(\frac{\omega L}{c}\right)}{\cos\left(\frac{\omega L}{c}\right)+igZ(L,\omega)\sin\left(\frac{\omega L}{c}\right)},
\label{eq:impedance}
\end{equation}
where $c=\sqrt{A_0(1-F_j)/\rho C}$ is wave propagation velocity. Similar to the large vessels, junction conditions impose pressure continuity and mass conservation. 

The terminal impedance of each structured tree is assumed to be zero for all frequencies, and impedance is calculated through bifurcations recursively to compute the impedance at the root of each structured tree \cite{Olufsen2001}. 

\vspace{0.2 cm}

\noindent\textbf{Wave intensity analysis.} \hot{Wave intensity is approximated using the pressure and flow predictions from the fluids model.} The propagated pressure and flow waves are decomposed using wave-intensity analysis (WIA), a tool that quantifies the incident and reflected components of these waves \cite{Kim,LaPole2022,Qureshi2019}. Components are extracted from the pressure and average velocity using
\begin{equation}
\label{eq:wia1}
    \Gamma_{\pm} (t) = \Gamma_0+\int_0^Td\Gamma_{\pm}, \hspace{5mm} \Gamma=p,\bar{u}
\end{equation}
\begin{equation}
\label{eq:wia2}
    dp_{\pm} = \frac{1}{2}(dp\pm\rho c\,d\bar{u}), \hspace{3mm} d\bar{u}_{\pm}=\frac{1}{2}\left(d\bar{u}\pm\frac{dp}{\rho c}\right),
\end{equation}
where $c$ (cm/s) is the pulse wave velocity, $dp$ the change in pressure and $d\bar{u}$ the change in average velocity across a wave \cite{Kim}. The time-normalized wave intensity is given by
\begin{equation}
    WI_{\pm} = \left(\frac{\text{d}p_\pm}{\text{d}t}\right)\left(\frac{\text{d}\bar{u}_\pm}{\text{d}t}\right),
\end{equation}
where the positive subscript refers to the incident wave and the negative to the reflective wave. Both incident and reflected waves can be classified as compressive or expansive. Compressive waves occur when $WI_\pm,\text{d}p_\pm>0$, while expansive waves occur when $WI_\pm,\text{d}p_\pm<0$. A wave-reflection coefficient is calculated by defining the ratio of amplitudes of reflected compressive pressure waves to incident compressive pressure waves
\begin{equation}
    I_R = \frac{\Delta p_-}{\Delta p_+}.
\end{equation}
The incident (forward) wave begins at the network inlet. Forward \textbf{compression} waves (FCW) increase pressure and flow velocity as blood travels downstream to the outlet. Forward \textbf{expansion} waves (FEW) decrease pressure and flow velocity. Reflective (backward) waves begin at the outlet of the vessel. Backward \textbf{compression} waves (BCW) increase pressure and decrease flow velocity as blood is reflected upstream of the vessel while backward \textbf{expansion} waves (BEW) decrease pressure and accelerate flow \cite{Colebank21,LaPole2022}.

\vspace{0.2 cm}

\noindent\textbf{Wall shear stress.} The stress, denoted$\tau_w$ (g/cm/s$^2$), the fluid exerts on the vessel wall, called the wall shear stress (WSS), is defined using the Stokes boundary layer from equation \eqref{eq:stokes}:
\begin{align*}
    \tau_w&= -\mu\frac{\partial u_x}{\partial r}\\
   &= \left\{
  \begin{array}{lr} 
      0, \hspace{1cm} & r<R-\delta, \\
      \displaystyle\frac{\mu\bar{u}}{\delta}, \hspace{1cm} & R-\delta<r\leq R, 
      \end{array}
      \right.
\end{align*}
where $\mu$ is kinematic viscosity, $\delta$ the thickness of the boundary layer, and $\bar{u}=q/A$ the average velocity \cite{Bartolo2022}.

\subsection{Model summary}
\label{sec:summary}
The 1D CFD model, described in section \ref{sec:Fluids}, predicts pressure, area, and flow in all 17 large vessels within the network shown in Figure \ref{fig:network}. Solutions of the model rely on parameters that specify the fluid and vessel properties. The latter includes geometric parameters that determine the vessel lengths, radii, connectivity, material parameters that determine vessel stiffness, and boundary condition parameters that determine inflow at the root of the network and outflow conditions at the terminal vessels. The model consists of both large vessels modeled explicitly and small vessels described by the structured tree framework.

\vspace{0.2 cm}

\noindent\textbf{Fluid dynamics.} Parameters required to specify the fluid dynamics  include blood density $\rho$ (cm$^3$), dynamic $\nu$ (g/cm/s) and kinematic $\mu = \nu/\rho$ (g/cm/s) viscosities, and the boundary layer thickness $\delta = \sqrt{\nu T/2\pi}$ (cm).
\[
\boldsymbol\theta_{\text{fluid}} = \{\rho,\mu,\nu,\delta\}.
\]
Hematocrit and viscosity are assumed to be the same for all patients, and, therefore, we keep these fixed for all simulations. Parameters $\rho, \mu$, and $\nu$ are used in the large and small vessels. Standard values for these parameters (Table \ref{tab:nomparam}) are obtained from the literature \cite{Olufsen2001}.

\vspace{0.2 cm}

\noindent\textbf{Geometry parameters.} For each patient, geometric parameters are determined by segmenting the MRA images as described in section \ref{sec:Network}. The network includes $17$ vessels, so there are $3\times17$ parameters corresponding to vessel lengths ($L$), inlet radii $r_\text{in}$, and outlet $r_\text{out}$ radii
\[
\boldsymbol\theta_{\text{geometry}} = \{r_{\text{in},i},r_{\text{out},i},L_i\}, \ \ \ i = 1,..17.
\]
\hot{These are the unstressed vessel dimensions.} Values for these parameters (Table \ref{table:Dim}) are extracted from the imaging data and are kept constant for each patient.  

\begin{table*}[t]
\centering
\caption{Vessel dimensions. Values denote the average of each measurement $\pm$ one standard deviation. Vessel numbers correspond to those shown in Figure \ref{fig:network}. Exact dimensions for individual patients can be found in the supplement. }
\label{table:Dim}  
\adjustbox{max width = \textwidth}{
\begin{tabular}{ll|rll|rll}
\hline\noalign{\smallskip}
& & \multicolumn{3}{|c|}{\textbf{DORV}} & \multicolumn{3}{|c}{\textbf{HLHS}}\\
\noalign{\smallskip}\hline\noalign{\smallskip}
 \textbf{Number} & \textbf{Name} & $L$ \textbf{(cm)} & $R\text{in}$ \textbf{(cm)} & $R_\text{out}$ \textbf{(cm)} & $L$ \textbf{(cm)} & $R_\text{in}$ \textbf{(cm)} & $R_\text{out}$ \textbf{(cm)} \\
\noalign{\smallskip}\hline\noalign{\smallskip}
1 & Asc aorta & $3.8\pm0.8$ & $1.18\pm0.08$ & $1.04\pm0.03$ & $2.9\pm0.9$ & $1.36\pm0.25$ & $1.52\pm0.22$ \\
2 & AA I & $1.6\pm0.6$ & $1.04\pm0.03$ & $1.00\pm0.04$ & $1.5\pm1.0$ & $1.52\pm0.22$ & $1.31\pm0.28$\\
3 & Innom & $4.3\pm1.2$ & $0.53\pm0.21$ & $0.53\pm0.21$ & $3.1\pm1.1$ & $0.54\pm0.12$ & $0.54\pm0.12$ \\
4 & AA II & $2.5\pm0.9$ & $1.00\pm0.04$ & $0.87\pm0.14$ & $1.5\pm0.5$ & $1.31\pm0.28$ & $0.74\pm0.18$ \\
5 & LCC & $20.1\pm0.7$ & $0.32\pm0.03$ & $0.31\pm0.04$ & $20.1\pm0.2$ & $0.3\pm0.05$ & $0.27\pm0.03$\\
6,10 & Subcl & $3.3\pm0.1$ & $0.36\pm0.09$ & $0.34\pm0.11$& $3.5\pm0.4$ & $0.21\pm0.04$ & $0.28\pm0.02$\\
7 & Desc aorta & $35.1\pm0.5$ & $0.87\pm0.14$ & $0.59\pm0.02$ & $35.1\pm0.2$ & $0.74\pm0.18$ & $0.55\pm0.02$\\
8 & L Brachial & $20.1\pm0.7$ & $0.31\pm0.05$ & $0.26\pm0.05$ &  $20.1\pm0.2$ & $0.30\pm0.05$ & $0.28\pm0.02$\\
9,17 & Vertebral & $14.3\pm0.5$ & $0.24\pm0.01$ & $0.23\pm0.03$ & $14.3\pm0.2$ & $0.24\pm0.03$ & $0.22\pm0.03$\\
11 & RCC & $17.1\pm0.6$ & $0.32\pm0.07$ & $0.30\pm0.07$ & $17.1\pm0.2$ & $0.34\pm0.08$ & $0.31\pm0.06$\\
12,15 & Ext C & $17.1\pm0.6$ & $0.27\pm0.08$ & $0.24\pm0.11$ & $17.1\pm0.2$ & $0.28\pm0.06$ & $0.26\pm0.07$\\
13,14 & Int C & $17.0\pm0.6$ & $0.21\pm0.05$ & $0.17\pm0.05$ & $17.0\pm0.2$ & $0.24\pm0.05$ & $0.23\pm0.06$ \\
16 & R Brachial & $20.1\pm0.7$ & $0.21\pm0.03$ & $0.21\pm0.03$ & $20.1\pm0.2$ & $0.24\pm0.03$ & $0.23\pm0.04$ \\
\noalign{\smallskip}\hline
\end{tabular}}
\end{table*}

\vspace{0.2 cm}

\noindent\textbf{Material parameters.} The model assumes vessels have increasing stiffness with decreasing radii. This trend is modeled by equation (\ref{eq:stiff}), which contains three stiffness parameters: $k_1$ (g/cm/s$^2$), $k_2$ (1/cm), and $k_3$ (g/cm/s$^2$). Vessels are grouped by similarity, and we keep these parameters fixed within each group. Referring to Figure \ref{fig:network}, vessels 1, 2, 4, and 7 (the aortic vessels) are in group 1, Vessels 3, 6, 8, 10, and 16 (vessels leading to the arms) are in group 2, vessels 5, 11, 12, 13, 14, and 15 (carotid vessels) are in group 3, and vessels 9 and 17 (vertebral vessels) are in group 4. In summary, we have 12 material parameters for the large vessels:
\[
\boldsymbol\theta_{\text{material}}=\{k_{1,g},k_{2,g},k_{3,g}\},  
\]
where subscript $g = 1, ..., 4$ enumerates the vessel groups. In our previous study \cite{LaPole2022}, these parameter values were determined using hand-tuning for a single DORV and HLHS patient pair. \hot{For this study, we use parameter values from \cite{LaPole2022} for $k_{1,g}, k_{2,g}$, while nominal values for $k_{3,g}$  are obtained by equations (\ref{eq:wallmodel}) and (\ref{eq:stiff})}. Nominal parameter values are listed in Table \ref{tab:nomparam}.

\vspace{0.2 cm}

\noindent\textbf{Boundary condition parameters.} The last set of parameters correspond to either the inflow or the structured tree boundary conditions. For each patient, at the inlet of the network, an inflow waveform is extracted from the 4D-MRI image. For the outflow boundary conditions, seven structured tree parameters are needed for each terminal vessel (vessels 7, 8, 9, 12, 13, 14, 15, 16, and 17). Parameters include the radius scaling factors $\alpha$ and $\beta$, that govern the asymmetry of the structured tree, $lrr$, that specifies the length-to-radius ratio, and $r_{\text{min}}$, the minimum radius used to determine the depth of the structured tree. In addition, the small vessels also have three stiffness parameters, $ks_1, ks_2,$ and $ks_3$. Small vessel stiffness parameters are vessel specific: vessel 7, the descending aorta, is in group 1, vessels 8 and 16, the brachial arteries, are in group 2, vessels 12 to 15, the carotid arteries, are in group 3, and vessels 9 and 17, the vertebral vessels, are in group 4. This gives a total of 16 structured tree parameters:
\[
\boldsymbol\theta_{\text{boundary}}=\{ks_{1,g}, ks_{2,g},ks_{3,g},\alpha,\beta,lrr,r_{\text{min}}\}.
\]
Nominal values for $\alpha, \beta, lrr, r_{\text{min}}, ks_{1,g},$ and $ks_{2,g}$ are taken from \cite{LaPole2022}, and $ks_{3,g}$ is calculated in the same way as $k_{3,g}$. For all terminal vessels except the descending aorta, the parameter $r_{\text{min}}$ is set to $0.001$ cm. This value is consistent with the average radius of the arterioles. The parameter $r_{\text{min}}$ for the descending aorta is set to $0.01$ cm, since this vessel is terminated at a larger radius.

\vspace{0.2 cm}

\noindent\textbf{Quantity of interest.} We estimate identifiable unknown parameters to quantify differences between the two patient groups. Our quantity of interest measures the discrepancy between model predictions and available data. Figure \ref{fig:registration} marks locations for flow waveform measurements extracted from the ascending aorta, descending aorta, and the innominate, left common carotid, and left subclavian arteries. Flow measured in the ascending aorta is used as the inflow boundary condition, and the remaining four flow waveforms are used to calibrate the model. We also have systolic and diastolic pressure measurements from the brachial artery. These are not measured simultaneously with the flows but are obtained in the supine position. Using this data, we construct residual vectors for both the flow and pressure. For each flow $q_i$ we compute
\begin{equation}
    r_{q_i}(t_j) = \left[\frac{q_i^m(t_{j})-q_i^d(t_j)}{q_{i,\max}^d}\right], \hspace{0.2cm} i=1,\dots 4, \hspace{0.2cm} j=1,\dots T_p \label{eq:res_q},
\end{equation}
where $r_{q_i}(t_j)$ denotes the residuals between the flow data ($q_i^d(t_j)$ (mL/s)) and the associated model predictions ($q_i^m(t_j)$ (mL/s)) at each of the four locations ($i=1,\dots4$) at time $t_j$ ($j = 1, \ldots T_p$), denoting the $j^{\textrm{th}}$ time step within the cardiac cycle. $T_p$ refers to the number of time point measurements (19-28) per cardiac cycle. This number differs between patients. The combined flow residual is defined as
\begin{equation}
    {\bf r_{q}} = [{\bf r}_{{\bf q}_1} \, {\bf r}_{{\bf q}_2} \, {\bf r}_{{\bf q}_3} \,{\bf r}_{{\bf q}_4}],
\label{eq:resq}
\end{equation}
where ${\bf r}_{{\bf q}_i}$ is the residual vector for flow $i$ defined in (\ref{eq:res_q}). It should be noted that $\mathbf{r_q}$ has dimensions of $4\times T_p$. 

Pressure measurements are available from one location at peak systole and diastole; therefore, the pressure residual is defined as
\begin{equation}
    \mathbf{r}_p = [r_{p_1}, r_{p_2}] = \left[\left(\frac{p_\text{sys}^m-p^d_\text{sys}}{p^d_\text{sys}}\right),\left(\frac{p^m_\text{dia}-p^d_\text{dia}}{p^d_\text{dia}}\right)\right],
    \label{eq:res_p}
\end{equation}
where $p_i^d$ and $p_i^m$, $i=\text{sys}, \text{dia}$ (mmHg) denote the data and model predictions respectively. Since we do not have the exact location of the cuff measurement, we match model predictions to data at the midpoint of the brachial artery.  
We perform sensitivity and identifiability analysis to determine a subset of influential and identifiable parameters. The initial set of parameters to be explored include those not determined from the patient data, i.e.,~the material and structured tree parameters,
\begin{equation}
\boldsymbol{\theta}_{\text{SA}} = \{k_{1,g},k_{2,g},k_{3,g},ks_{1,g},ks_{2,g},ks_{3,g},\alpha,\beta,lrr\}.
\label{eq:thetasens}
\end{equation}

\begin{table*}
\centering
\caption{Nominal parameter values for each patient group. {\bf D} refers to the DORV group and {\bf H} to the HLHS group. Parameters selected for inference are bolded. $r_{min}$ varies among the terminal vessels, so it is listed with a range.}
\label{tab:nomparam}
    \begin{tabular}{l l r r r }
    \hline\noalign{\smallskip}
    \textbf{Parameter} & \textbf{Description} & \textbf{Unit} & \textbf{Value (D)} & \textbf{Value (H)}\\
    \hline\noalign{\smallskip}
    $\rho$ & Density & g/cm$^3$ & 1.06 & 1.06\\
    $\mu$ & Constant viscosity & g/cm-s & 0.032 & 0.032 \\
    $\nu$ & Kinematic viscosity & cm$^2$/s & 0.030 & 0.030\\
    $\delta$ & Boundary layer thickness & cm & $\sqrt{\nu T/2\pi}$ & $\sqrt{\nu T/2\pi}$\\
    $T$ & Cardiac cycle length & s & $0.710\pm0.090$ & $0.770\pm0.190$\\
    $k_1$ & Large vessel stiffness & g/cm-s$^2$ & $2.0\times10^7$ & $2.0\times10^7$\\
    $k_2$ & Large vessel stiffness & cm$^{-1}$ & $-25.0$  & $-35.0$ \\
    $\mathbf{k_{3,g}}$ & \textbf{Large vessel stiffness} & \textbf{g/cm-s}$^2$ & $\mathbf{5.6\pm1.1\times10^5}$ & $\mathbf{7.3\pm2.1\times10^5}$\\
    $ks_1$ & Small vessel stiffness & g/cm-s$^2$ & $2.0\times10^7$ & $2.0\times10^7$\\
    $\mathbf{ks_2}$ & \textbf{Small vessel stiffness} & \textbf{cm}$^{-1}$ & $\mathbf{-35.0}$ & $\mathbf{-30.0}$\\
    $\mathbf{ks_{3,g}}$ & \textbf{Small vessel stiffness} & \textbf{g/cm-s}$^2$ & $\mathbf{3.8\pm2.2\times10^5}$ & $\mathbf{5.4\pm2.1\times10^5}$\\
    $\boldsymbol{\alpha}$ & \textbf{ST asymmetry constant} & Non.Dim. & $\mathbf{0.900}$ & $\mathbf{0.90-}$\\
    $\beta$ & ST asymmetry constant & Non.Dim. & 0.600 & 0.600 \\
    $lrr$ & Length to radius ratio & Non.Dim & 50.0 & 50.0 \\
    $r_{min}$ & Minimum radius & cm & $0.001$-$0.010$ & $0.001$-$0.010$\\
    \hline\noalign{\smallskip}
\end{tabular}
\end{table*}

\subsection{Sensitivity and identifiability analysis}

Local sensitivity analysis provides insight into the influence of nominal parameters on the model predictions. However, sensitivities can be inaccurate if the model is highly nonlinear and optimal parameters vary significantly from their nominal values. Global sensitivities provide additional information in the form of parameter influences over the entire parameter space. Local and global sensitivity analyses are performed on the flow residual vector ${\bf r_{q}}$ defined in (\ref{eq:resq}) with respect to parameters $\boldsymbol{\theta}_{\text{SA}}$ defined in (\ref{eq:thetasens}).

\vspace{0.2 cm}

\noindent\textbf{Local sensitivity analysis.} Using a derivative-based approach, we compute local sensitivities of the quantity of interest (residual vector) with respect to each parameter. Sensitivities are evaluated by varying one parameter and fixing all others at their nominal values \cite{Colebank_sens}. We compute local sensitivities to log-scaled parameters $(\Tilde{\boldsymbol{\theta}}_{\text{SA}}=\log\boldsymbol{\theta}_{\text{SA}})$, ensuring both positivity and that parameters values are on the same scale. With these assumptions, the local sensitivity of $\bold{r_{q}}$, with respect to the $n^{\textrm{th}}$ component of the parameter vector, is given by
\begin{align*}
    \mathbf{S}_{n} &= \frac{\partial \bold{r_{q}}}{\partial\Tilde{\theta}^n_{\text{SA}}}\\
    &= \left[\frac{\partial\mathbf{q}_1}{\partial\theta^n_{\text{SA}}}\frac{\Tilde{\theta}^n}{q_{1,\max}^d},\dots,
    \frac{\partial\mathbf{q}_4}{\partial\theta^n_{\text{SA}}}
    \frac{\Tilde{\theta}^n}{q_{4,\max}^d}\right], \hspace{0.2cm} n=1\dots B,   
\end{align*}
where $B$ denotes the total number of parameters. Sensitivities are estimated using centered finite differences
\begin{equation*}
    \pard{\bold{r_{q}}}{\Tilde{\theta}^n_{\text{SA}}} \approx \frac{\bold{r_{q}}(\Tilde{\theta}^n_{\text{SA}}+\mathbf{e}_n\psi)-\bold{r_{q}}(\Tilde{\theta}^n_{\text{SA}}-\mathbf{e}_n\psi)}{2\psi},  
\end{equation*} 
where $\tilde{\theta}^n_{\text{SA}}$ is the parameter of interest, $\psi$ is the step size, and $\mathbf{e}_n$ is a unit basis vector in the $n^\textrm{th}$ direction \cite{Colebank_sens,pope2009}. Parameters are ranked from  most to least influential by calculating the 2-norm of each sensitivity \cite{Colebank_sens,Colunga23},
\begin{equation}
    \bar{\mathbf{S}}_n = ||\tilde{\mathbf{S}}_n||_2.
\end{equation}

\noindent\textbf{Global sensitivities.} Morris screening \cite{Sumner2012,Colebank_sens,MorrisScreening} is used to compute global sensitivities. This method predicts elementary effects defined as
\begin{equation}
    \mathbf{d}_n(\theta^n_{\text{SA}}) = \frac{\mathbf{r_q}(\theta^n_{\text{SA}}+\mathbf{e}_n\Delta)-\mathbf{r_q}(\theta^n_{\text{SA}})}{\Delta},
\end{equation}
where the number of samples is set to $K=100$ and the number of levels of parameter space is set to $\mathcal{M}=60$, resulting in a step size of $\Delta = \frac{\mathcal{M}}{2(\mathcal{M}-1)} \approx 0.508$.  
Elementary effects are determined by sampling $K$ values from a uniform distribution for a particular parameter $\theta^n_{\text{SA}}$. The elementary effects are ranked by computing the 2-norm of each effect,
\[\Tilde{d}_n^j(\theta) = ||d_n^j(\theta)||_2,\]
where $j$ denotes the time point. Using the algorithm by Wentworth \textit{et al.} \cite{MorrisScreening}, results are integrated to determine the mean and variance for the elementary effects. These are defined as
\begin{align}
    \mu_n^* &= \frac{1}{K}\sum_{j=1}^K|\Tilde{d}_n^j|,\\
    \sigma^2_n &= \frac{1}{K-1}\sum_{j=1}^K\left(\Tilde{d}_n^j -\mu_n^*\right)^2,
\end{align}
where $\mu^*$ is the sensitivity of the quantity of interest and $\sigma^2$ is the variability in sensitivities due to parameter interactions and model nonlinearities. Parameters are ranked by computing $\sqrt{\mu^{*2}+\sigma^2}$ to account for the magnitude and variability of each elementary effect. To stay within a physiological range of parameter values, large and small vessel stiffness parameters are perturbed $\pm10\%$, and $\alpha,\beta,lrr$ are perturbed $\pm5\%$. 

\vspace{0.2 cm}

\noindent\textbf{Covariance analysis.} Pairwise correlations between parameters can be determined using covariance analysis, constructing a covariance matrix,
\begin{equation}
    v_{nj} = \frac{V_{nj}}{\sqrt{V_{nn}V_{jj}}},
\end{equation}
where $V = s^2[\Tilde{\mathbf{S}}_n^T\Tilde{\mathbf{S}}_n]^{-1}$ and $s^2$ is a constant observation variance. Studies using this method have defined correlated parameters for which $v_{nj}>0.8$ to $0.95$ \cite{Olsen2018,Marqui2018,Colebank_sens,Brady2018}.  We assume that parameter pairs are correlated if $v_{nj}>0.9$. Information gained from the sensitivity and covariance analyses allows us to determine a parameter set for inference, denoted $\boldsymbol{\theta}_{\text{inf}}$, which, as we will show in Section~\ref{ssec:ParamInferenceResults}, is given by
\begin{equation}\label{eq:InferenceParVector}
    \boldsymbol{\theta}_{\text{inf}} = \{\alpha, k_{3,1:4},ks_2, ks_{3,1:4}\}.
\end{equation}

\subsection{Parameter inference}
The parameter subset $\boldsymbol{\theta}_{\text{inf}}$ defined in equation \eqref{eq:InferenceParVector} is inferred by minimizing the RSS
\begin{equation}
   RSS = \sum_{i=1}^4 \left( \sum_{j=1}^{T_p} r_{q_i}(t_j)^2 \right) + \sum_{k\in \{\text{dia}, \text{sys}\}} r_{p_k}^2,
   \label{eq:cost}
\end{equation}
where $r_{q_i}(t_j)$ and $r_{p_k}$ are defined in equations~\eqref{eq:res_q} and \eqref{eq:res_p}. To minimize equation (\ref{eq:cost}), we use the sequential quadratic programming (SQP) method, a \hot{second-order} gradient-based algorithm  \cite{Colebank_sens,Boggs95}, that is implemented in Matlab's \texttt{fmincon} function with option 'sqp'. \hot{Several previous studies have succesfully used this algorithm on similar problems \cite{Qureshi2019,Paun2018,Colebank19}.} We use a tolerance of $1.0\times 10^{-8}$ \cite{matlab}. Parameter bounds for each patient are set to ensure that model predictions remain within the physiological range. Exact nominal values and ranges for each patient are listed in the supplemental material. 

To complement the global sensitivity analysis, we use multistart inference to test if the locally identifiable parameter subset remains identifiable. The optimizer is initialized to $12$ sets of parameter values specified by sampling from a uniform distribution defined by varying the parameters $\pm30\%$ of their nominal values. We record the final value of the cost function and check for convergence across optimizations. Parameters with a coefficient of variation (CoV, the standard deviation divided by the mean) greater than 0.10 are removed from the subset. The multistart inference is repeated until all parameters in the subset converge and each the coefficient of variation for each parameter is below $0.10$. Parameters in the final subset, collected in the vector  $\boldsymbol{\theta}_{\text{inf}}$, are estimated through one round of optimization with different initializations to avoid trapping in local minima.

\section{Results}
\subsection{Image analysis}
The DORV and HLHS patients have significantly different vessel radii. Figure \ref{fig:radii} shows that remodeling of the reconstructed aorta in the HLHS group widens the ascending aorta and aortic arch as compared to the native aorta in the DORV group. For most HLHS patients, the aortic arch is wider than the ascending aorta. However, the two groups have approximately the same radii at the distal end of the descending aorta. 

\vspace{0.2 cm}
\noindent\begin{minipage}{\columnwidth}
\captionsetup{type=figure}
\centering\includegraphics[width=0.8\textwidth]{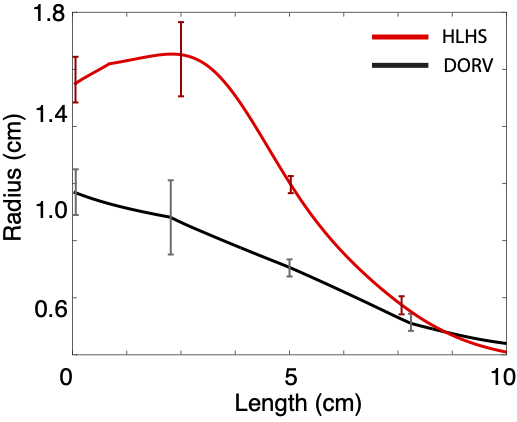}
    \caption{Aortic radii for a DORV and an HLHS patient pair. In the HLHS patient, the radii increase along the ascending aorta and then decrease, while in the DORV patient, the aorta decreases along its length. Note that the two vessels have similar radii at the distal end. Error bars (one standard deviation of radii measurements) are plotted at the inlet of the ascending aorta, aortic arch I, aortic arch II, and descending aorta. Only the beginning of the descending aorta differs between the two patients.}
    \label{fig:radii}
\end{minipage}
\vspace{0.2 cm}

The geometric results in Table \ref{table:Dim} show that remodeling is heterogeneous. The variance of vessel dimensions (see supplement for details) along the ascending aorta and aortic arch aorta is significantly smaller in the DORV group compared to the HLHS group, while the variance within each patient is similar (Figure \ref{fig:radii}). Remodeling impacts the vessels in contact with the reconstructed tissue, namely the innominate artery and aortic arch. The aortic arch is widened, and the innominate artery is significantly shorter in HLHS patients. The same holds for the ascending aorta, which is also shorter in HLHS patients. The head and neck vessels and the descending aorta have similar dimensions and variance across groups, but as noted in Section \ref{sec:modelpred}, these vessels are stiffer in HLHS patients. For each patient type, vessel dimension ranges are listed in Table \ref{table:Dim}, and exact vessel dimensions are reported in the supplement.

\subsection{Parameter inference}\label{ssec:ParamInferenceResults}

\noindent\textbf{Parameter identifiability.} An identifiable parameter subset is constructed by combining the results of the local and global sensitivity analyses, covariance analysis, and multistart inference. 

\begin{figure*}
    \centering
    \includegraphics[width=1.0\textwidth]{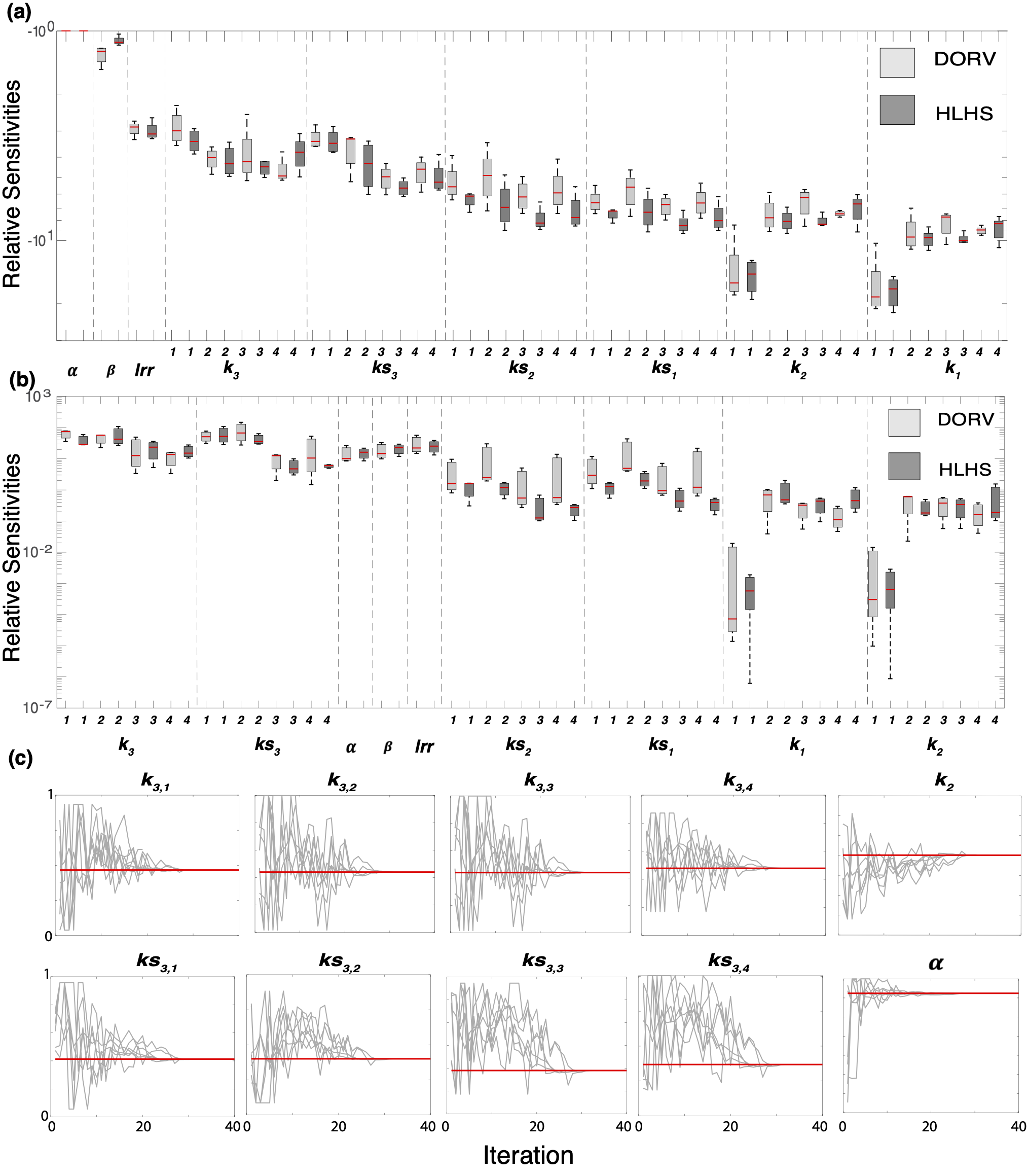}
    \caption{Sensitivity analysis and multistart inference. \textbf{(a)} local sensitivities, \textbf{(b)} Morris screening, and \textbf{(c)} Multistart inference on the final identifiable parameter subset. Panels \textbf{(a)} and \textbf{(b)} are boxplots of the relative sensitivities for each patient group. Sensitivities are ranked from most to least influential. Panel \textbf{(c)} shows convergence of the final identifiable parameter subset. For each parameter, CoV $< 0.1$. Note that parameters are scaled from 0 to 1 for easier comparison.}
    \label{fig:SA}
\end{figure*}

Local sensitivity analysis reveals that the most influential parameters are $\alpha$, $\beta$, and $lrr$. Other influential parameters include vessel stiffness parameters, $k_{3,g}$, $ks_{3,g}$, and $ks_{2,g}$ (see Figure \ref{fig:SA}). Globally, $k_{3,g}$ and $ks_{3,g}$ are the most influential, followed by the structured tree parameters $\alpha, \beta, lrr$. However, the relative sensitivities for these five parameters are similar, i.e., they are good candidates for the parameter subset. This result should be contrasted with parameters $k_{1,g}$ and $k_{2,g}$, which are less influential. 

Covariance analysis revealed a correlation between the structured tree parameters. Since $\alpha$ is the most influential parameter, we inferred $\alpha$ and fixed $\beta$ and $lrr$. The large vessels stiffness parameters $k_{3,g}$ and $k_{1,g}$ are also correlated, and so are $k_{2,g}$ and $k_{1,g}$. Informed by the sensitivity analyses, we chose to fix $k_{1,g}$ and $k_{2,g}$ and inferred only $k_{3,g}$. For small vessels, $k_{2,g}$ is more influential, therefore we fixed $ks_1$ and inferred $ks_{2,g}$ and $ks_{3,g}$.

The initial parameter subset includes parameters $\theta_{\text{CA}}=\{\alpha, k_{3,1:4}, ks_{2,1:4}, ks_{3,1:4}\}$ that are influential and locally  uncorrelated. Multistart inference of $\theta_{\text{CA}}$ resulted in Cov > 0.1 for $ks_{2,g}$. Given that most structured trees are similar, we set $ks_{2,g}= ks_2$. Repeated estimation including $ks_2$ as a global parameter gave an identifiable subset including $\boldsymbol \theta_{\text{inf}} =\{\alpha, k_{3,1:4},ks_2, ks_{3,1:4}\}$, i.e., all 12 samples estimated parameters with a CoV < 0.1. Results of the multistart are shown in Figure \ref{fig:SA}(c).

\noindent

\vspace{0.2 cm}

\begin{figure*}[b]
    \centering
    \includegraphics[width=1.0\textwidth]{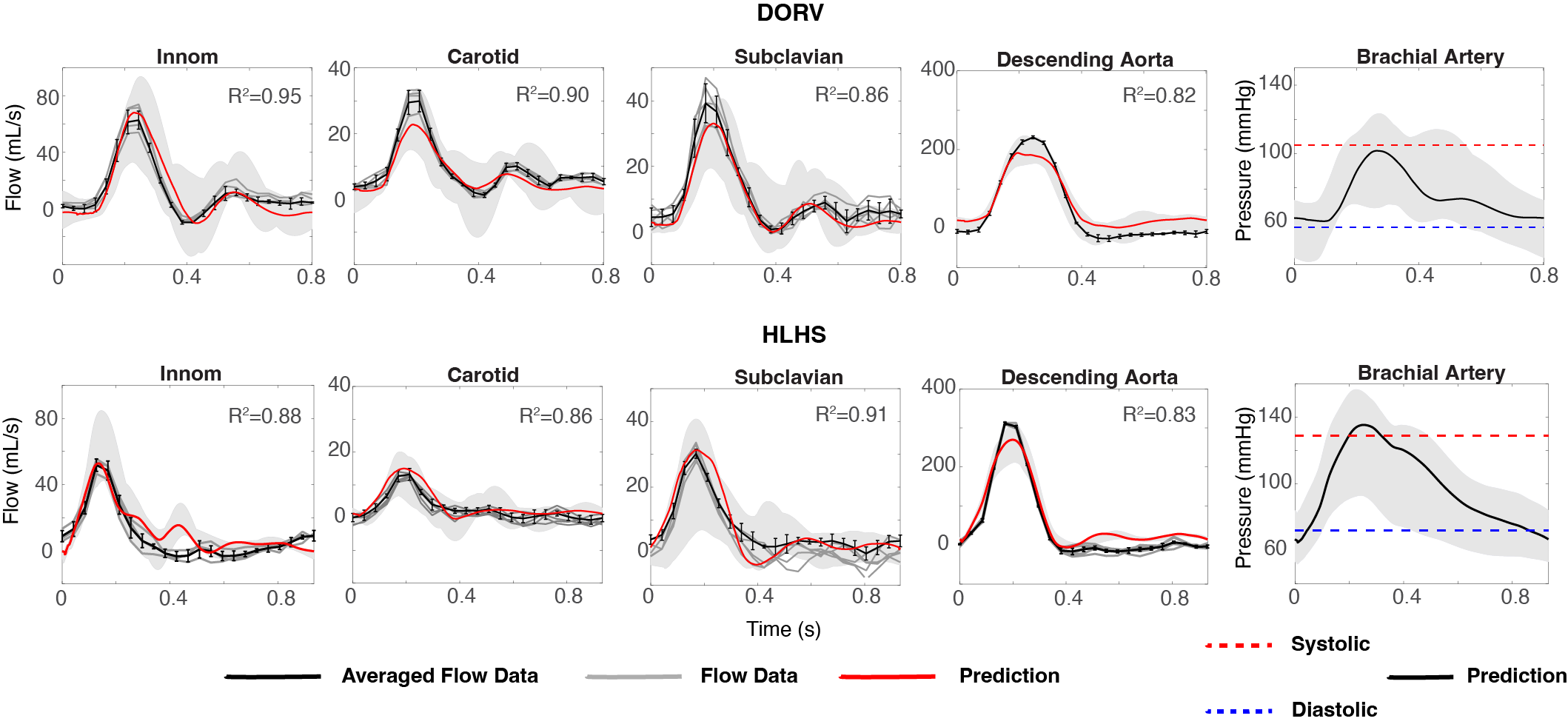}
    \caption{Model predictions versus patient data. From left to right, the first four plots in each row depict flow waveforms. The five gray lines denote dynamics for each patient and the the averaged dynamics is plotted with a black line. The black error bars represent one standard deviation of the averaged waveform. Model predictions are shown in red. The far-right plots in each row show the systolic (dashed red) and diastolic (dashed blue) cuff pressure measurements. Model predictions are depicted using solid black lines. Shading in the background of each plot denote the results of simulations sampling the optimized parameters from a uniform distribution. \hot{$R^2$ values reporting the quality of model predictions is given in the top right of each plot.}}
    \label{fig:DatavsPred1}
\end{figure*}

\noindent\textbf{Model calibration.}  Parameter inference results shown in Figure~\ref{fig:DatavsPred1} reveal that the model fits both the main and reflected features of the flow waveforms and the systolic and diastolic brachial pressures. Optimal model predictions are plotted with a solid red line and the averaged measured flow waveforms with solid black lines. Error bars were obtained by averaging waveforms extracted at nearby points within the vessel, as described in Section \ref{sec:Data}. These bars correspond to one standard deviation above and below the mean. The gray silhouette represents predictions generated by sampling parameters from a uniform distribution within the bounds used for parameter inference, demonstrating the ability of the optimization to generate model outputs that fit the data. \hot{For flow waveforms, $R^2$ values were always $>0.80$, indicating the model captures at least $80\%$ of variability in the data \cite{Ratner09}}.

\begin{figure*}
    \centering
    \includegraphics[width=6.5in]{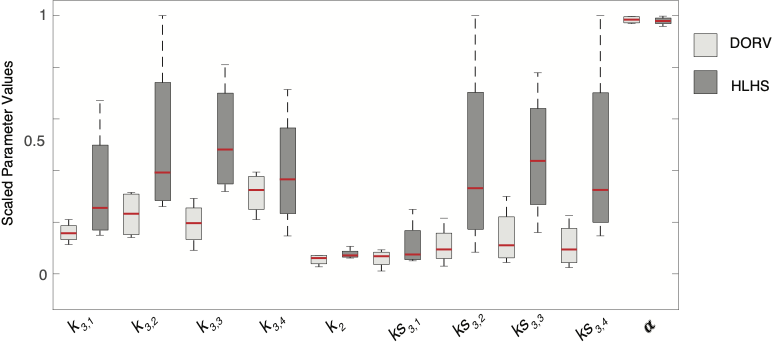}
    \caption{Box and whisker comparison of inferred parameter values. Light gray boxes represent the DORV patients, while dark gray represent the HLHS group. Parameters are scaled from $0$ to $1$ for easier comparison.}
    \label{fig:boxplotparam}
\end{figure*}

\subsection{Model predictions}
\label{sec:modelpred}
\noindent\textbf{Inferred parameters.} Figure \ref{fig:boxplotparam} compares the estimated parameters, scaled from $0$ to $1$. These results show that the aortic stiffness ($k_{3,1}$) is higher and more variable in HLHS patients, which appears to be consistent with our results for the vessel geometry. Even though vessels further from the reconstruction have similar geometry between patient groups, vessel stiffness ($k_{3,2:4}$) is increased in HLHS, indicating remodeling affects the vasculature as a whole. Moreover, peripheral vessels are stiffer in HLHS patients, but remodeling has not affected the peripheral branching structure (estimated values for $\alpha$ are the same for all vessels). This result seems to agree with the finding that geometry is not altered in the peripheral vessels.

\vspace{0.2 cm}

\begin{figure*}[t]
    \centering
    \includegraphics[width=1.0\textwidth]{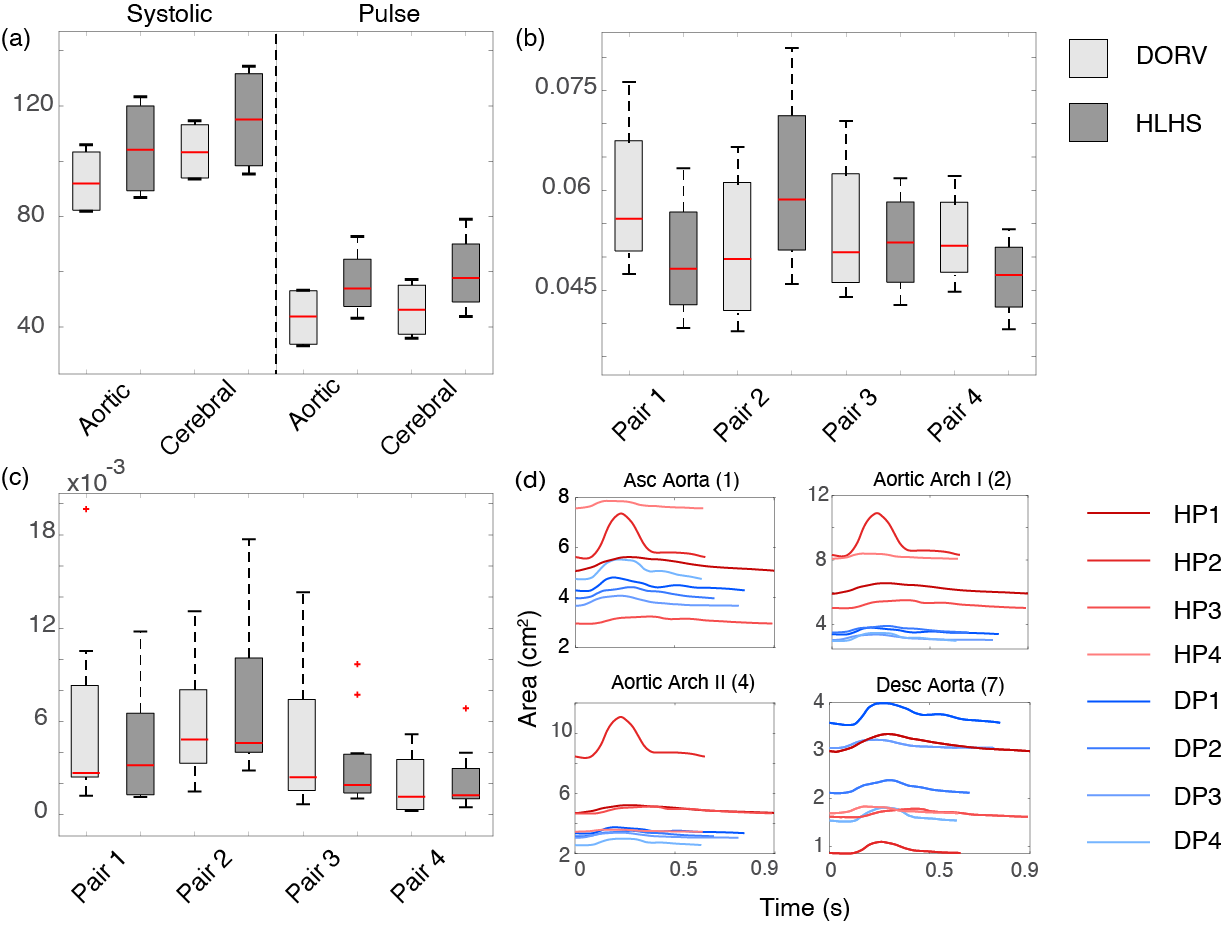}
    \caption{Model predictions. (a) The average systolic and pulse pressure in the aorta and cerebral vessels. Pairwise comparison of the average pulse flow in the aorta (b) and cerebral (c) vessels. Pulse flows are shown relative to total cardiac output.\hot{(d) Area deformation over a cardiac cycle, HLHS patients are shown in red and DORV in blue.}}
    \label{fig:pressure_area}
\end{figure*}

\noindent\textbf{Flow, pressure, and area predictions.}  Figure \ref{fig:pressure_area}(a) shows that average systolic and pulse pressures are higher in the aortic and cerebral arteries for the HLHS patients. Figure \ref{fig:pressure_area} (b) and (c) shows pairwise comparisons of flow predictions. Averaging of all patients within each group shows no differences, but pairwise comparisons of pulse flow (max flow - min flow) reveal different trends. In (b), pulse flow is lower in HLHS patients or has no significant difference in aortic vessels, except in HLHS patient 2 (from patient pair 2). In (c), on average, pulse flow is decreased in the HLHS cerebral vasculature. This can be seen better by comparing the values vessel-wise (see supplement). Paired HLHS and DORV patients have similar cardiac outputs, therefore this pairwise comparison provides greater insight. Larger pulse flows in cerebral vessels indicates greater perfusion to the brain, while similar pulse flows in the descending aorta suggests both groups are possibly at the same risk for FALD (see supplement). Note that the reported pulse flows are relative to the total cardiac output. \hot{Figure \ref{fig:pressure_area} (d) shows that HLHS patients' aortic vessels deform less over a cardiac cycle compared to DORV.}
\noindent

\begin{figure*}
    \centering
    \includegraphics[width=1.0\textwidth]{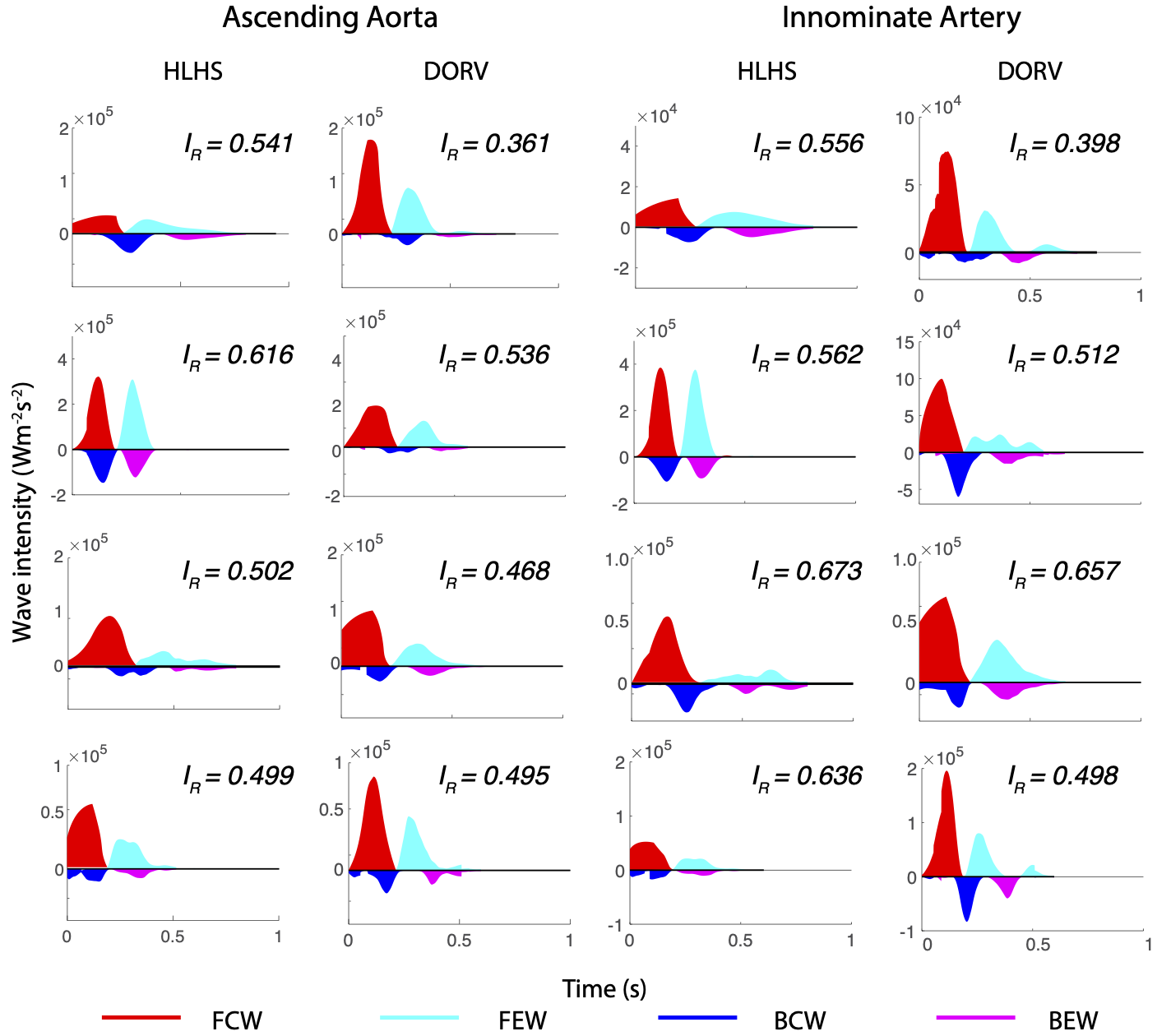}
    \caption{Reflected backward compression waves (BCW) and backward expansion waves (BEW). Incident forward compression waves (FCW) and forward expansion waves (FEW)) waves in the ascending aorta and left/right carotid artery for each patient. The patterns in the the other aortic and peripheral vessel segments are similar to the vessels shown here. Each panel also lists the patient's wave-reflection coefficient, $I_R$.}
    \label{fig:WIA}
\end{figure*}
\vspace{0.2 cm}

\noindent\textbf{Wave intensity analysis.} The incident and reflected waves shown in Figure \ref{fig:WIA} differ significantly between the two patient groups. Overall, the wave-reflection coefficient $I_r$ is higher in HLHS patients, suggesting reconstructed aortas induce more reflections. This might be in part due to larger differences in vessel size and stiffer vessels for patients in the HLHS group. Moreover, DORV patients have significantly higher forward compression and expansion waves and HLHS patients have higher backward compression waves, likely a result of the widened ascending aorta and aortic arch, increasing the wave reflections. The exception to this trend is pair number 2. The DORV patients do not differ significantly from the other patients in the group, but HLHS patient 2 has significantly higher forward and backward waves. This patient does not have significantly different geometry as compared to the other patients. However, they do have increased pulse flow in their aortic vessels. Our modeling approach can predict abnormalities in this patient by taking in to account the complex interactions of these quantities. 

\vspace{0.2 cm}

\begin{figure*}[t]
    \centering
    \includegraphics[width=1.0\textwidth]{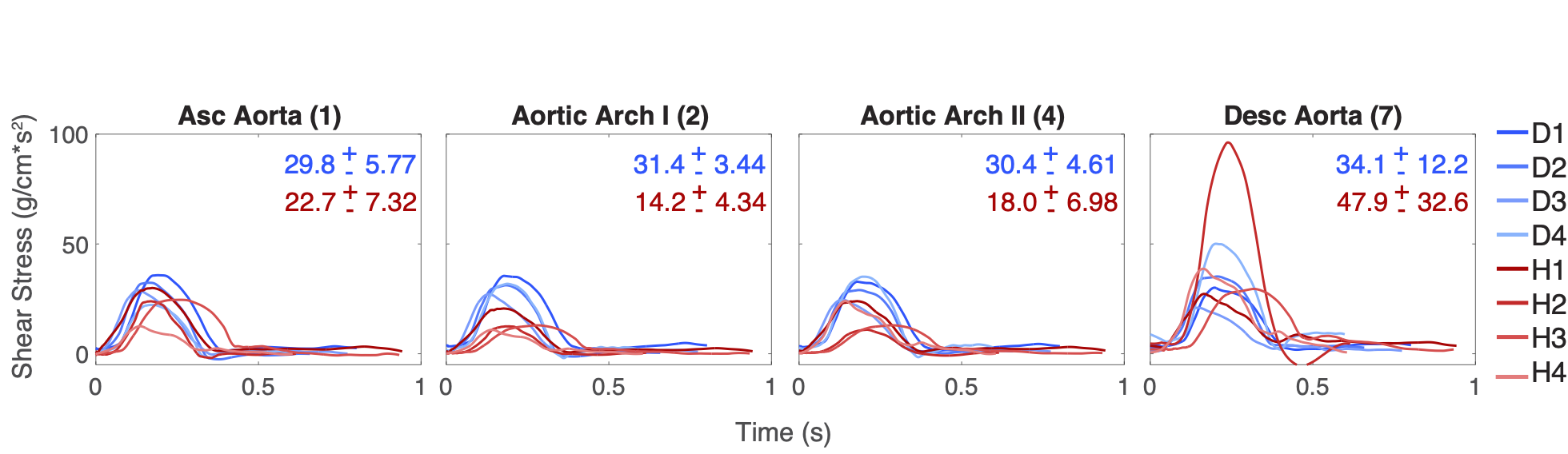}
    \caption{Wall shear stress in the four aortic vessels for each patient. Shades of blue represent DORV patients, and shades of red represent HLHS patients. The average $\pm$ one standard deviation peak WSS for the  DORV and HLHS groups is noted on the graphs. DORV patients' WSS decreases slightly away from the heart, while HLHS patients' WSS increases. For all patients, the WSS is significantly higher in the descending artery. HLHS patient 2 has considerably higher descending aorta WSS than the other patients.}
    \label{fig:WSS}
\end{figure*}

\noindent\textbf{Wall shear stress.} We computed wall shear stress (WSS) in the center of the aortic vessel segment for each patient (Figure \ref{fig:WSS}). DORV patients have higher WSS values that peak between 25 to 50 g/cm$\cdot$s$^2$ compared to HLHS patients, with WSS values peaking between 10 to 29 g/cm$\cdot$s$^2$. All patients have similar WSS values in the ascending aorta and aortic arch, but it is significantly higher in the DORV patients in these regions.  Both groups have increased wall shear stress in the descending aorta, but the increase is higher in HLHS patients. This is likely a result of the stiffening of the descending aorta in this patient group. Again, HLHS patient 2 has larger WSS in the descending aorta compared to the other patients.

\section{Discussion}
This study describes a framework for building patient-specific 1D CFD models and  applies these models to predict hemodynamics for DORV and HLHS single ventricle patients. We analyze medical images to extract patient geometries and to quantify differences in vessel dimensions; perform local and global sensitivity analyses, covariance analysis, and multistart inference to determine an influential and identifiable subset of parameters. We infer this subset, $\theta_{\text{inf}}=\{k_{3,g},ks_2,ks_{3,g},\alpha\}$. Results show that HLHS patients have wider ascending aortas, increased vascular stiffness throughout the network, increased pressures in the aorta and cerebral vasculature, and increased wave reflections. Additionally, model predictions reveal that HLHS patient 2 has quantitatively different pressure and flow patterns compared to the other HLHS patients. 

\subsection{Image analysis}
Patient geometry derived from medical imaging data provide known parameters that is integrated into the model. It has a substantial influence on the model predictions. Colebank \textit{et al.} \cite{Colebank19} and Bartolo \textit{et al.} \cite{BartoloLaPole2023} detail the importance of accurate vascular geometry and its influence on model predictions. HLHS patients have greater ascending aorta radii. Two HLHS patients have significant remodeling, with ascending aortas increasing along the aortic arch rather than decreasing. However, all patients have similarly sized descending aortas. This abnormal geometry inherent to the HLHS cohort likely contributes to abnormal flow patterns and disturbances, promoting continued vascular and ventricular remodeling over time \cite{Chiu2013,Renna2013,Hayama2020}. Hayama \textit{et al.} \cite{Hayama2020} noted that aortic roots in HLHS patients continue to dilate over time, contributing to both ventricular and downstream remodeling and eventual Fontan failure. Therefore, monitoring the change of aortic geometry over time is essential for single ventricle patients with reconstructed aortas.

\subsection{Parameter inference}
Local sensitivity analysis reveals that the structured tree parameters $\alpha$ and $\beta$ have the greatest influence on the quantity of interest. Stiffness parameters $k_{3,g}$ and $ks_{3,g}$ also influence the predictions significantly. This result agrees with findings by Paun \textit{et al.} \cite{SECRET} who used a similar 1D CFD model. Their study only considered local sensitivities and parameters were not vessel specific. They concluded that $\alpha$ was the most influential, followed by $\beta$ and $k_3$. Our findings also agree with work from Clipp and Steele \cite{Clipp2009}, which emphasized the importance in tuning of the structured tree boundary condition parameters to fit the model to data. 

Morris screening results are consistent with local sensitivity results in the sense that the five most influential parameters are the same in each case. Figure \ref{fig:SA} shows that while $k_{3,g}$ and $ks_{3,g}$ are slightly more influential than the structured tree parameters, there are no significant differences in the values of the relative sensitivities. Our findings are also consistent with experimental studies investigating the effects of aortic stiffness and resistance on hemodynamics. In the aortic vasculature, it has been found that increased aortic stiffness can decrease stroke volume and increase pressure \cite{London1999,Gulan2014}. Through an in-vitro investigation focused on the impact of aortic stiffness on blood flow, Gulan \textit{et al.} \cite{Gulan2014} found that increasing aortic stiffness greatly influences velocity patterns and blood volume through the aorta. 

Covariance analysis revealed a correlation between the structured tree parameters, as well as between $k_{1,g}$ and the stiffness parameters $k_{2,g}$ and $k_{3,g}$. We inferred $\alpha$ and fixed $\beta$ and $lrr$. The parameter $\alpha$ is the most influential, and it is not correlated with any of the stiffness parameters. We fixed $k_1$ and $k_2$ for all vessels to reduce the number of inferred parameters. Covariance analysis is performed using the local sensitivity matrix, therefore there is no guarantee that parameters remain correlated after being estimated. Multistart inference results showed that $ks_{2,g}$ had a CoV > 0.1, and estimated parameters varied significantly. Due to the influence that $ks_2$ has on the model, we inferred that parameter but used a global rather than vessel-specific value. These findings are consistent with studies from Colebank \textit{et al.} \cite{Colebank2023} and Paun \textit{et al.} \cite{SECRET} in which they performed related  analyses on similar models and chose to infer large artery stiffness, small vessel stiffness, and at least one structured tree parameter.

Figure \ref{fig:boxplotparam} compares the inferred parameter values between the two patient groups. Overall, DORV patients have lower large vessel stuffiness in the aorta and peripheral vessels. The stiffness varies less between vessel types, and hemodynamic predictions are more uniform. Estimated values for the $\alpha$ and $ks_2$ parameters did not differ significantly between groups. However, small and large vessel stiffness, $ks_{3,g}$ and $k_3$ respectively, were substantially higher in the HLHS group, with DORV patients having lower downstream resistance vascular stiffness than HLHS patients. This finding is consistent with those from Cardis \textit{et al.} \cite{Cardis2006} and Schafer \textit{et al.} \cite{Schafer2019}, who found that Fontan patients with HLHS had higher vascular stiffness compared to other Fontan patient types. The increased stiffness might occur in part due to properties of the non-native tissue used to surgically reconstruct the aorta \cite{Cardis2006}. For most HLHS patients, reconstruction is performed with a homograft material that comprises at least $50\%$ of the reconstructed vessel \cite{Cardis2006}. This homograft material differs significantly from the native aortic tissue, and remodeling over time generates tissue that is significantly  stiffer than the native aorta 
\cite{Azadani2011,Jia2017}. On a related note, clinicians have discussed the utility of aortic stiffness in helping to determine when medical intervention is needed. The retrospective study by Hayama \textit{et al.} \cite{Hayama2020} found that increased aortic stiffness in Fontan patients is correlated with exercise intolerance, vascular and ventricular remodeling, and heart failure. They postulated that surgical intervention and vasodilation/hypertension medication may help offset vascular remodeling \cite{Hayama2020}.

\subsection{Model predictions}
\noindent{\bf Hemodynamic predictions.} Figure \ref{fig:DatavsPred1} demonstrates that our parameter inference method generates model predictions that fit the data reasonably well. We found that the HLHS group has increased average systolic and pulse pressures with the interquartile range (IQR) also being higher in aortic and cerebral vasculature. Given the heterogeneous geometry present in the HLHS group and the small sample size, the latter is only evident with comparisons between patient pairs. Since HLHS requires surgical interventions over multiple years, we found that using matched DORV patients provided a clear and systematic way to understand the effects of surgical aortic reconstruction and subsequent remodeling on model predicted hemodynamics. All HLHS patients are hypertensive \cite{Blanco2017} according to model predictions, despite three out of four HLHS patients receiving medication to reduce blood pressure. In particular, cerebral blood pressure was high. Decreased pulse flows are consistent with research that describes inadequate perfusion and oxygen transport to the brain in HLHS patients with reconstructed aortas \cite{Navaratnam2016,Schneider1914,pulseflow}. Brain perfusion appears to be an important clinical endpoint, since it has recently been shown that HLHS patients have abnormal cerebral microstructure and delayed intrauterine brain growth \cite{Mahle2004,Oberhuber2017}. With respect to vessel area deformation over a cardiac cycle, the aortic vessels in the HLHS patients (except for HLHS patient 2) deform less than the same vessels in the DORV patients (see supplement). Vessels with small deformations over a cardiac cycle tend to have increased stiffness and tend to appear in patients with larger aortic radii and abnormal wall properties \cite{Biglino2012}.

\vspace{0.2 cm}

\noindent{\bf Wave intensity analysis.} Figure \ref{fig:WIA} shows DORV have larger forward waves than their reflective, but HLHS have smaller forward waves and increased reflective waves. In HLHS patients, the ascending aorta has smaller forward waves and larger backward waves compared to the DORV group. Wave-reflection coefficients $I_R$ (Figure \ref{fig:WIA}) confirm these differences. For each matched patient pair, the HLHS patient has a larger $I_R$. The $I_R$ values we found for DORV patients' agree with those reported in literature \cite{Pomella2018}. The results of our WIA for the ascending aorta of the HLHS group are consistent with a similar analysis by Schafer \textit{et al.} \cite{Schafer2021}, who noted increased ratios of backward to forward waves in HLHS patients with reconstructed aortas. \hot{Increased backward compression waves in the ascending aorta indicate a sudden deceleration of the forward blood flow wave, leading to an increase in afterload and a decrease in ventricular performance \cite{Schafer2021}.} In particular, HLHS patient 2 has a significantly higher $I_R$ in the aortic vessels compared to the other patients. 
\hot{These results signify abnormal blood flow within the aorta of HLHS patients, possibly indicating ventricular decline.}

\vspace{0.2 cm}

\noindent{\bf Wall shear stress.} 
Except for DORV patient 3, WSS peak values in the DORV group decrease in the descending aorta compared to the ascending aorta. This finding is consistent with studies using CFD models to compute WSS in healthy patients \cite{Karmonik2013,Callaghan2018}. The HLHS group has lower WSS values in the ascending aorta and aortic arch compared to the DORV group. Reduced WSS can be indicative of hypertension and stiffening of the vessel wall.  Traub \textit{et al.} \cite{Traub1998} found that consistently low WSS values were correlated with upregulation of vasoconstrictive genes. This promoted smooth muscle cell growth, which led to a loss of vessel compliance. For the larger WSS values in the descending aorta, Voges \textit{et al.} \cite{Voges2015} found that the descending aorta, in HLHS patients with reconstructed aortas, dilate over time due to increased WSS. Notably, HLHS patient 2 has a large increase in WSS in the descending aorta. Of the HLHS group, this patient has the smallest inlet radius for the descending aorta. As the body begins to remodel, this WSS could decrease as the inlet radius widens. WSS is an important clinical marker that cannot be measured in vivo. However, it can be determined from CFD models, as demonstrated by Loke \textit{et al.} \cite{Loke2020}. They used CFD modeling and surgeon input to develop a Fontan conduit that minimized power loss and shear stress, thereby improving flow from the gut to the pulmonary circuit. \hot{Many studies have focused on WSS in the Fontan conduit and pulmonary arteries \cite{Rijnberg2021,Visser2008}. However, there is a lack of work devoted to studying WSS within the aorta and systemic arterial vasculature for single ventricle patients. The results reported here give crucial insight into hemodynamics and information on the degree of remodeling.}

\subsection{Future work and limitations}
This study describes the construction of patient-specific, 1D CFD arterial network models that include the aorta and head/neck vessels for four DORV and four HLHS patients. A main contribution is the development of a parameter inference methodology that uses multiple datasets. To obtain reliable parameter inference results, we limited the number of vessels in the network. The small size of the network makes it challenging to predict cerebral and gut perfusion. A way to overcome this limitation is to add more vessels, e.g., to use the network defined in the study by Taylor-LaPole {\em et al.} \cite{LaPole2022}. Another limitation is that HLHS patients generally have abnormal aortic geometries due to surgical reconstruction. These geometries most likely cause energy losses as blood flows from the wider arch into the narrow descending aorta. Our model does not predict these energy losses. However, previous studies have included energy loss terms in 1D arterial network models \cite{Colebank21,Mynard2015}. This approach could be adapted for this study, but more work is needed to calibrate parameters required for these energy loss models. Calibration could be informed by the analysis of velocity patterns from 4D-MRI images or 3D fluid-structure interaction models. 

\hot{The Fontan circuit has been the subject of many studies using three-dimensional (3D) CFD models \cite{Marsden2007,Marsden2009,Ahmed2021}. While these models are excellent for analyzing complex velocity patterns, they are limited to the imaged region and are generally computationally intensive. In contrast, 1D  CFD models offer an efficient and accurate alternative. Several studies have shown that the flow and pressure waves predicted by 1D models are comparable to those obtained with 3D models \cite{Moore2005,Reymond2012,Blanco2018}, validating the use of 1D models. Future research will explore optimization techniques that allow this model to be used in real-time clinical decision-making, a possibility made feasible by the efficiency of the 1D model over 3D modeling. Future work will explore optimization techniques that allow this model to be used in real-time clinical decision making - made possible by the efficiency of the 1D model over 3D modeling.}

This study minimizes the Euclidean distance between measurements and model predictions inferring biophysical parameters of interest, ignoring the correlation structure of the measurement errors due to the temporal nature of the data. In future studies, we aim to capture the correlation by assuming a full covariance matrix for the errors by using Gaussian Processes \cite{Paun2020}. We will also incorporate model mismatch to account for discrepancies between data and model predictions (Figure~\ref{fig:DatavsPred1}) caused by numerical errors or model assumptions by using the methodology presented in \cite{Paun2020}. \hot{We did not include analyses of high-order interaction from multiple parameters being considered, this will be explored in future work. Also, our sensitivity analyses were performed with respect to $r_q$ as our data consisted mainly of flow waveforms. On one patient we did perform these same analyses including $r_p$ (data not shown) and this method was slightly less efficient and did not produce different results.} Finally, many modeling studies devoted to the Fontan circulation focus only on venous hemodynamics, in part because venous congestion impacts blood returning to the heart from the liver and likely contributes to the progression of FALD. In the future, a two-sided vessel network model could be incorporated into this framework that includes a description of both the arterial and venous vasculature, see e.g.~\cite{Puelz2017}.

There are limitations related to the clinical data that was used in this study. The 4D-MRI images are averaged over several cardiac cycles. This averaging and noise in the scans likely contributes to a lack of volume conservation in the flow data created from these images. We performed parameter inference using one pressure reading from one vessel. In the future, it would be helpful for parameter estimation to have multiple pressure readings from multiple vessels, e.g., from the arms and the ankles. \hot{We do realize the sample size is small - limiting our ability to validate our model. It is our plan in the future to obtain access to more patient data and validate results in this way}. Finally, five out of the nine patients received some form of hypertension medication. This is a factor that our model does not take into account. 

\section{Conclusions}
This study defines a patient-specific 1D CFD model and a parameter inference methodology to calibrate the model to 4D-MRI velocity data and sphygmomanometer pressure data. Results from the parameter inference give insight into physiological phenomena such as vascular stiffness and downstream resistance. The reconstructed aortas in the HLHS patients were wider than the native aortas of the DORV patients, and parameter inference revealed that HLHS patients has increased vascular stiffness and downstream resistance. Model predictions showed that vessels in HLHS patients do not distend over a cardiac cycle as much as those in DORV patients, indicative of hypertension. WIA predicted increased backward waves in the ascending aortas of HLHS patients, suggesting abnormal blood flow. Results shows decreased WSS in HLHS patients indicative of hypertension and a precursor to remodeling. HLHS patient 2 in particular has the highest pressures, largest backward waves, and largest WSS of the HLHS patient group, indicating this patient may be in need of additional clinical, possibly surgical, intervention. To our knowledge, this study is the first patient-specific 1D CFD model of the Fontan systemic arterial vasculature that is calibrated using multiple data sets from multiple patients.

\enlargethispage{20pt}

\dataccess{Software and code for the fluids model and optimization can be found at github.com/msolufse.}

\aucontribute{AMT designed the study, performed all analyses and hemodynamic simulations, and drafted the manuscript. DL performed image registration. DL, JDW, and CP provided all patient data and advised its use and contributed to writing and editing the manuscript. LMP provided parameter inference/statistical expertise and contributed to writing and editing the manuscript. MSO conceived and coordinated the study and helped write and edit the manuscript.}

\competing{We have no competing interests.}

\funding{This work was supported by the National Science Foundation (grant numbers DGE-2137100, DMS-2051010). Any opinions, findings, and conclusions expressed in this material are those of the authors and do not necessarily reflect the views of the NSF. Work carried out by LMP was funded by EPSRC, grant reference number EP/T017899/1.}

\ack{We thank Yaqi Li for segmenting the patient geometries for the image registration.}

%%%%%%%%%% Insert bibliography here %%%%%%%%%%%%%%

\printbibliography

\end{multicols}

\end{document}